\documentclass[a4paper, 12pt]{article}
\usepackage[margin=2cm]{geometry}
\usepackage{times, amssymb, amsmath, graphicx, tabularx, booktabs, subfiles, amsthm,enumitem,xcolor,braket, subcaption}

\usepackage{cite}

\newcommand{\td}[1]{\tilde{#1}}

\title{Rational Extension of Quantum Anisotropic Oscillator Potentials
with Linear and/or Quadratic Perturbations}
\author{
Rajesh Kumar Yadav$^{a}$\thanks{e-mail: rajeshastrophysics@gmail.com (R.K.Y.)},	Rajesh Kumar$^{a,b}$\thanks{e-mail: kr.rajesh.phy@gmail.com (R.K.)}, 
	and 
	Avinash Khare$^{c}$\thanks{e-mail: avinashkhare45@gmail.com (A.K.)}
}

\begin{document}
\maketitle

\begin{center}
	\text{$^a$Department of Physics, S. K. M. University, Dumka-814110, India} \\
	\text{$^b$Department of Physics, Model College, Dumka-814101, India} \\
	\text{$^c$Department of Physics, Savitribai Phule Pune University, Pune-411007, India} \\
\end{center}

\begin{abstract}
  We present a comprehensive study of the rational extension of the quantum  
anisotropic harmonic oscillator (QAHO) potentials with linear and/or quadratic 
perturbations. For the one-dimensional harmonic oscillator plus imaginary 
linear perturbation ($i\lambda x$), we show that the rational extension is 
possible not only for the even but also for the odd co-dimensions $m$. In 
two-dimensional case, we construct the rational extensions for QAHO potentials 
with quadratic ($\lambda \, xy$) perturbation both when $\lambda$ is real or 
imaginary and obtain their solutions. Finally, we extend the discussion to the 
three-dimensional QAHO with linear and quadratic perturbations and obtain the
corresponding rationally extended potentials. For all these cases, we obtain the
conditions under which the spectrum remains real and also when there is 
degeneracy in the system.

\end{abstract}

\section{Introduction}

After the discovery of the exceptional Jacobi and Laguerre orthogonal polynomials (EOPs) 
\cite{gomez2010extension, gomez2009extended}, many well known potentials have 
been rationally extended and their eigenfunctions have been obtained using  
these EOPs \cite{quesne2008exceptional, bqr, os, mi_os, mi_os2, yg11,rkaop13,
rkplb13,rkpla13,nkpt16,rkpara16, sbop22, sb24}. Later on, the rational 
extension of the one dimensional harmonic oscillator potential 
\cite{gomez2013rational} was also done and its solutions were obtained in terms
of the exceptional Hermite polynomials provided the co-dimension $m$ is even 
\cite{marquette2013two}. Subsequently, this work was extended using the 
supersymmetric quantum mechanics (SQM) approach. Recently, we obtained one 
parameter family of isospectral potentials as well as the uncertainty relations
corresponding to these $m$-dependent potentials \cite{kumar2023rationally}. 
Subsequently, the rational extension of the higher dimensional anisotropic 
harmonic oscillator (AHO) potentials has also been done \cite{Kumar:2024tsz}. 

It is then natural to enquire if one can obtain the rational extension of even 
more general AHO potentials. For example, can one obtain rational extension of 
the one dimensional harmonic oscillator along with linear perturbation or 
rational extension of the higher dimensional AHO potentials with linear and/or 
quadratic perturbations? The purpose of this paper is to answer some of the 
questions raised here. In particular, we show that unlike the one dimensional
harmonic oscillator (where rational extension is only possible for even 
co-dimension $m$), the rational extension of the one dimensional harmonic 
oscillator plus imaginary perturbation of the form $i\lambda x$ is possible not
only for the even but for the odd co-dimension $m$. Further, we show that the 
rational extension is also possible for the two dimensional AHO plus quadratic 
perturbation of the form $\lambda x y$. Generalization to higher dimensions is 
straight forward where one can also consider in addition either an admixture of
both the linear and the quadratic perturbation or a combination of the 
quadratic perturbations.

The key idea of our approach is to attack the problem in two steps. In the 
first step, we eliminate the linear and/or quadratic perturbation terms by 
transforming to new (tilde) coordinates, resulting in a pure AHO potential. 
These transformations include coordinate shifting in the case of the linear 
perturbation and coordinate rotation to absorb the quadratic perturbation. In 
the second step, we extend this pure AHO potential rationally, following the 
well-known procedure discussed in Refs. \cite{marquette2013two} and 
\cite{Kumar:2024tsz}. The resulting RE potential in the tilde coordinates is 
parameterized by $m$. Finally, by applying the inverse coordinate 
transformation, one obtains the rational extension of the original 
perturbed potential. Throughout these transformations, the Laplacian operator 
$\nabla^2$ remains invariant. 

The plan of the paper is as follows: In section $2$, we consider the rational 
extension of the one dimensional harmonic oscillator potential plus linear 
perturbation of the form $\lambda x$ on the full line. We show that when 
$\lambda$ is real, the rational extension is only possible for even 
co-dimension $m$. However, in case $\lambda$ is pure imaginary then unlike the 
pure oscillator, the rational extension is possible for both the even and the 
odd co-dimension $m$. In section $3$, we consider the rational extension of the
two dimensional AHO with linear and/or quadratic perturbations defined on the 
full line. In this case the rational extension is only possible for the even 
co-dimensions $m_1,m_2$. In Sec. IV we consider a three dimensional AHO along 
with (i) either a combinations of the linear and the quadratic perturbation 
(ii) or a combination of the quadratic perturbations and obtain the 
corresponding rational extension in both the cases. Finally in Section $5$, we 
summarize our findings and point out few possible open problems. A Brief 
description about the concept of the $\mathcal{PT}$ symmetry and the possible 
forms of the parity operators ($\mathcal{P}$) in higher dimensions are given in Appendix A.

\section{One-dimensional harmonic oscillator with linear perturbation}
Let us start with the well-known one-dimensional harmonic oscillator (QHO) 
potential $V(x)$ (defined on the full line i.e., $-\infty<x<\infty$) with 
\( \omega_1 \) as the angular frequency, satisfying the Schr\"odinger equation 
in the units of \(\hbar = 2m = 1\), given by

\begin{equation}\label{1D-SE}
\left[-\frac{d^2 }{dx^2} +V(x) \right] \psi(x) = E \psi(x);\quad V(x)=\frac{1}{4}\omega_1^2 x^2\,.
\end{equation}
The correspond eigenvalues and eigenfunctions are \cite{greiner2001quantum}
 \begin{equation}\label{QHO-psi}
\psi_n(x) \propto e^{-\frac{\omega_{1}}{4} x^{2}} 
	 H_n\left(\sqrt{\frac{\omega_1}{2}} x\right)
\end{equation}
and 
\begin{equation}\label{QHO-E}
 E_n(\omega_1)=\left(n+\frac{1}{2}\right)\omega_1;\quad n=0,1,2... \,.
\end{equation}
We now introduce a linear perturbation \( \lambda_0 x\) (with 
$\lambda_0$ real) given by
\begin{equation}\label{1D-V-I}
V^{l}(x) =V(x) + \lambda_0 x ,
\end{equation}
which can be reduced to the standard form of QHO (see for example 
\cite{greiner2001quantum}) by shifting the co-ordinate $x\rightarrow \td{x}$ 
and get
\begin{align}\label{1D-V'}
\td{V}^l(\td{x}) 
&= \frac{1}{4}\omega_{1}^{2} \td{x}^2 
-\frac{\lambda_0^2}{\omega_{1}^2},\quad\mbox{where}\quad \td{x} = x 
+ \frac{2\lambda_0}{ \omega_{1}^{2}}.
\end{align}
The eigenfunctions for the potential $\td{\psi}_n^l(\td{x})$ are
the same as those of QHO as given above except for a shift in the coordinate 
\( x \rightarrow \tilde{x} \).
If we replace $\lambda_0\rightarrow i\gamma_0$ (where $\gamma_0$ is real), the 
potential (\ref{1D-V-I}) is complex but $\mathcal{PT}$ invariant \cite{bender1998real,singh2019pt}. The 
corresponding energy eigenvalues are all real and given by 
\begin{equation}
E\rightarrow \tilde{E}_n^l(\omega_1)=\left(n+\frac{1}{2}\right) \omega_1
+\frac{\gamma^2_0}{\omega_1^2};\quad n=0,1,2...,
\end{equation} 
while the corresponding eigenfunctions are all $\mathcal{PT}$-invariant with $\mathcal{PT}$-eigenvalue
being $\pm1$ i.e.,
\begin{align}
    \begin{split}
    \mathcal{PT}\;\td{\psi}_n^l(\td{x})=(-1)^n\;\td{\psi}_n^l(\td{x}).
    \end{split}
\end{align}
\subsection{Rational extension: Real $\lambda_0$ case}
The rational extension of the unperturbed potential (\ref{1D-SE}) defined on 
the full-line can be easily obtained by using the results discussed in 
Refs. \cite{marquette2013two, Kumar:2024tsz} for even 
co-dimension\footnote{If we consider the potential defined on the half-line, 
the corresponding rational extension will be acceptable for all positive 
integer values of $m$ \cite{Kumar:2024tsz}. In this manuscript, we consider the
potential defined on the full-line only. The same procedure may also be adopted
to handle the potentials defined on the half-line.} of $m$. The expression for
the potential of the rationally extended quantum harmonic oscillator (RE-QHO) 
potential corresponding to the potential (\ref{1D-SE}) is given by  
\begin{equation}\label{1D-V-RE-x}
  V_{RE,m}(x) = V(x) + V_{rat,m}\left(x\right), \quad -\infty < x < \infty
\end{equation}
where the rational term $V_{rat,m}(x)$ is 
\begin{equation}\label{rat}
    V_{rat,m}(x)=-2\left[\frac{\mathcal{H}_m''\left(\sqrt{\frac{\omega_1}{2}} x\right)}{\mathcal{H}_m\left(\sqrt{\frac{\omega_1}{2}} x\right)} - \left[\frac{\mathcal{H}_m'\left(\sqrt{\frac{\omega_1}{2}} x\right)}{\mathcal{H}_m\left(\sqrt{\frac{\omega_1}{2}} x\right)}\right]^2 + \frac{\omega_1}{2}\right],\quad m=0,2,4,...
\end{equation}
The ground and the excited state eigenfunctions are 
\begin{eqnarray}\label{1D-psi-RE-x}
  \psi_{RE,m,0}(x) &\propto& \frac{\zeta\left(\omega_1,x\right)}{\mathcal{H}_m\left(\sqrt{\frac{\omega_1}{2}} x\right)}\nonumber\\
\mbox{and} \quad 
  \psi_{RE,m,n+1}(x) &\propto& \frac{\zeta\left(\omega_1,x\right)}{\mathcal{H}_m\left(\sqrt{\frac{\omega_1}{2}} x\right)}\hat{H}_{m,n+1}\left(\sqrt{\frac{\omega_1}{2}} x\right),
\end{eqnarray}
with  $\zeta(\omega_1,x)=e^{- \frac{\omega_1 }{4}x^2}$ respectively, where 
\( n = 0, 1, 2, \dots \) and \(\hat{H}_{m,0}(x) = 1\). Here 
$\hat{H}_{m,n+1}(x)$ is the exceptional Hermite polynomial 
\cite{marquette2013two} given by
\begin{align}
\hat{H}_{m,n+1}(x) &= \mathcal{H}_m(x) H_{n+1}(x) + H_n(x) \frac{d}{dx} \mathcal{H}_m(x),
\end{align}
where \( H_n(x) \) is the classical Hermite polynomial and 
\( \mathcal{H}_m(x) \) is the pseudo Hermite polynomial. The energy eigenvalues
\cite{Kumar:2024tsz} are given by
\begin{equation}\label{1D-E_RE-x}
E_{RE,m, n+1}(\omega_{1})=\left(n+m+1\right)\omega_1 \qquad\mbox{with}\quad E_{RE,m,0}(\omega_{1})=0.
\end{equation}
Therefore, the rational extension of the linearly perturbed potential 
(\ref{1D-V'}) in terms of the new coordinate $(\td{x})$ is obtained using the 
above results and is given as
\begin{equation}\label{1D-V-RE-I}
    \tilde{V}_{RE,m}^l(\td{x}) = \td{V}^l(\td{x}) + \td{V}^l_{rat,m}\left(\td{x}\right), \quad -\infty < x < \infty,
\end{equation}
where $\td{V}^l(\td{x})$ is the perturbed potential (\ref{1D-V'}) 
and $\td{V}^l_{rat,m}\left(\td{x}\right)$ is the corresponding rational part 
which can be simply obtained from Eq. (\ref{rat}) by replacing 
$x\rightarrow\td{x}$ i.e., 
\begin{equation}
\td{V}^l_{rat,m}\left(\td{x}\right)=V_{rat,m}\left(x\rightarrow\td{x}\right).
\end{equation}
The corresponding ground and the excited state eigenfunctions are given by
\begin{eqnarray}\label{1D-psi-RE-I}
	\td{\psi}_{RE,m,0}^l(\td{x}) &=& \frac{\zeta\left(\omega_1,\td{x}\right)}{\mathcal{H}_m\left(\sqrt{\frac{\omega_1}{2}} \td{x}\right)}\nonumber\\
	\mbox{and} \quad 
	\td{\psi}_{RE,m,n+1}^l(\td{x}) &=& \frac{\zeta\left(\omega_1,\td{x}\right)}{\mathcal{H}_m\left(\sqrt{\frac{\omega_1}{2}} \td{x}\right)}\hat{H}_{m,n+1}\left(\sqrt{\frac{\omega_1}{2}} \td{x}\right).
\end{eqnarray}
respectively. The energy eigenvalues of $\td{V}^l_{RE,m}(\td{x})$ are the same 
as $V_{RE,m}(x)$ as the factor of $\frac{\text{$\lambda $o}^2}{\omega_1^2}$ 
cancels out \cite{Kumar:2024tsz} and are given by
\begin{equation}\label{1D-E_RE-I}
\td{E}_{RE,m, n+1}^l(\omega_{1})=E_{RE,m,n+1} \qquad\mbox{with}\quad \td{E}_{RE,m,0}^l(\omega_{1})=E_{RE,m,0}=0.
\end{equation}
On using the expression for $\td{x}$ in the equation (\ref{1D-V-RE-I}), one can
easily obtain the RE potential $V_{RE,m}^l(x)$ (which is the rational extension
of the conventional potential $V^l(x)$) with the perturbation term. The 
explicit expressions for the RE potential $V_{RE,m}^l(x)$ and the
corresponding eigenfunctions ($\psi^l_{RE,m,n+1}(x)$) are shown in Tables 
\ref{Tab:1D-V-}, \ref{Tab:1D-psi0-} and \ref {Tab:1D-psin-} respectively in 
case $m = 0, 2$. It is worth pointing out that when \(\lambda_0\) is real, 
\(\tilde{x}\) is also real, and the potential becomes singular for odd \( m \) 
due to the presence of the \(\frac{1}{\tilde{x}^2}\) term, which diverges at 
the origin. However, in the case of even $m$, the potential has no such 
singularity.
 
\subsection {Rational extension: Imaginary $\lambda_0$ case}
 
\begin{figure}[htp]
  \centering
  \IfFileExists{complex-V.jpeg}{
      \includegraphics[height=0.55\textheight]{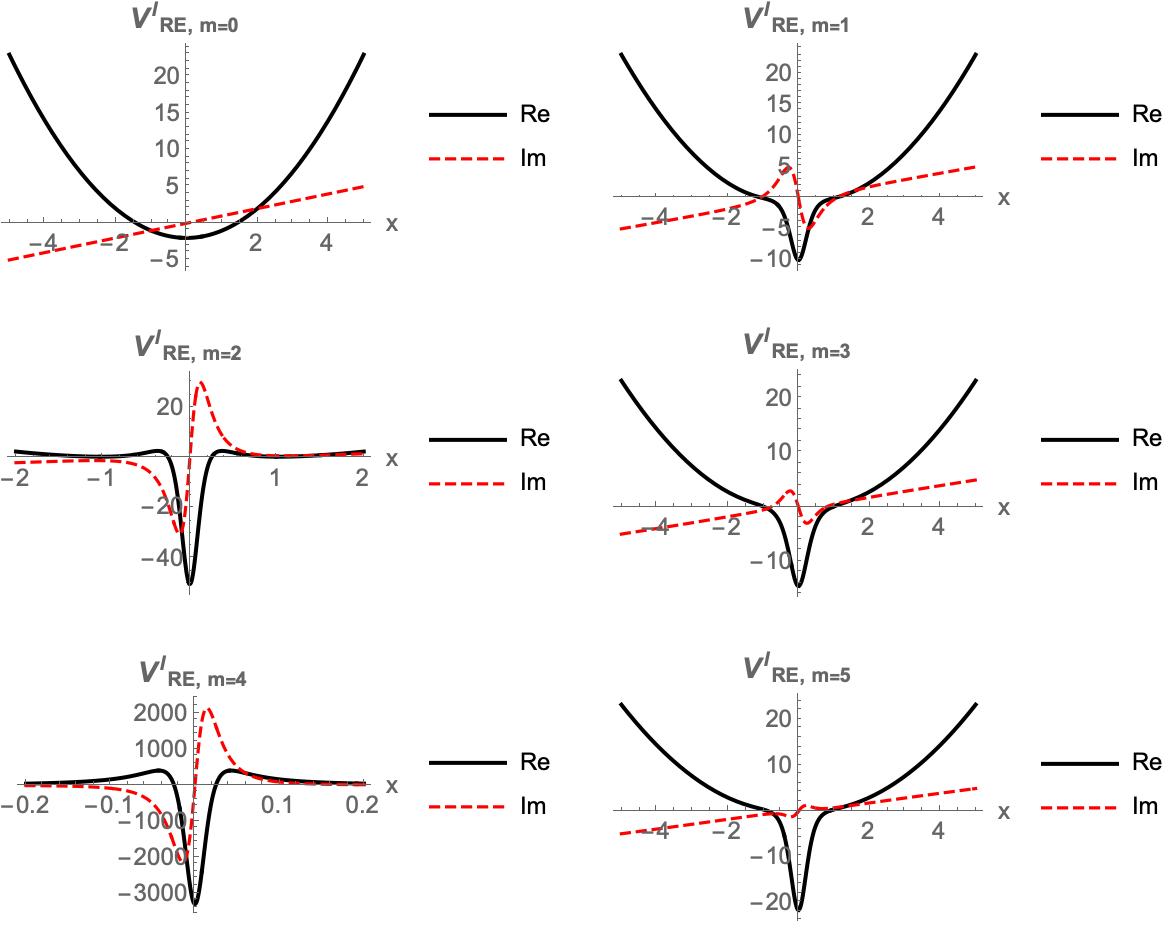}
  }{
      \fbox{\textbf{Figure Missing}}
  }
  \caption{Plots of real and imaginary components of $V^l_{RE,m}(x)$ vs $x$ for $m=0$ to $5$ with $\omega_1=2$.}
  \label{fig:1D-V-}
\end{figure}

\begin{figure}[htp]
  \centering
  \IfFileExists{Wave0.jpeg}{
      \includegraphics[height=0.55\textheight]{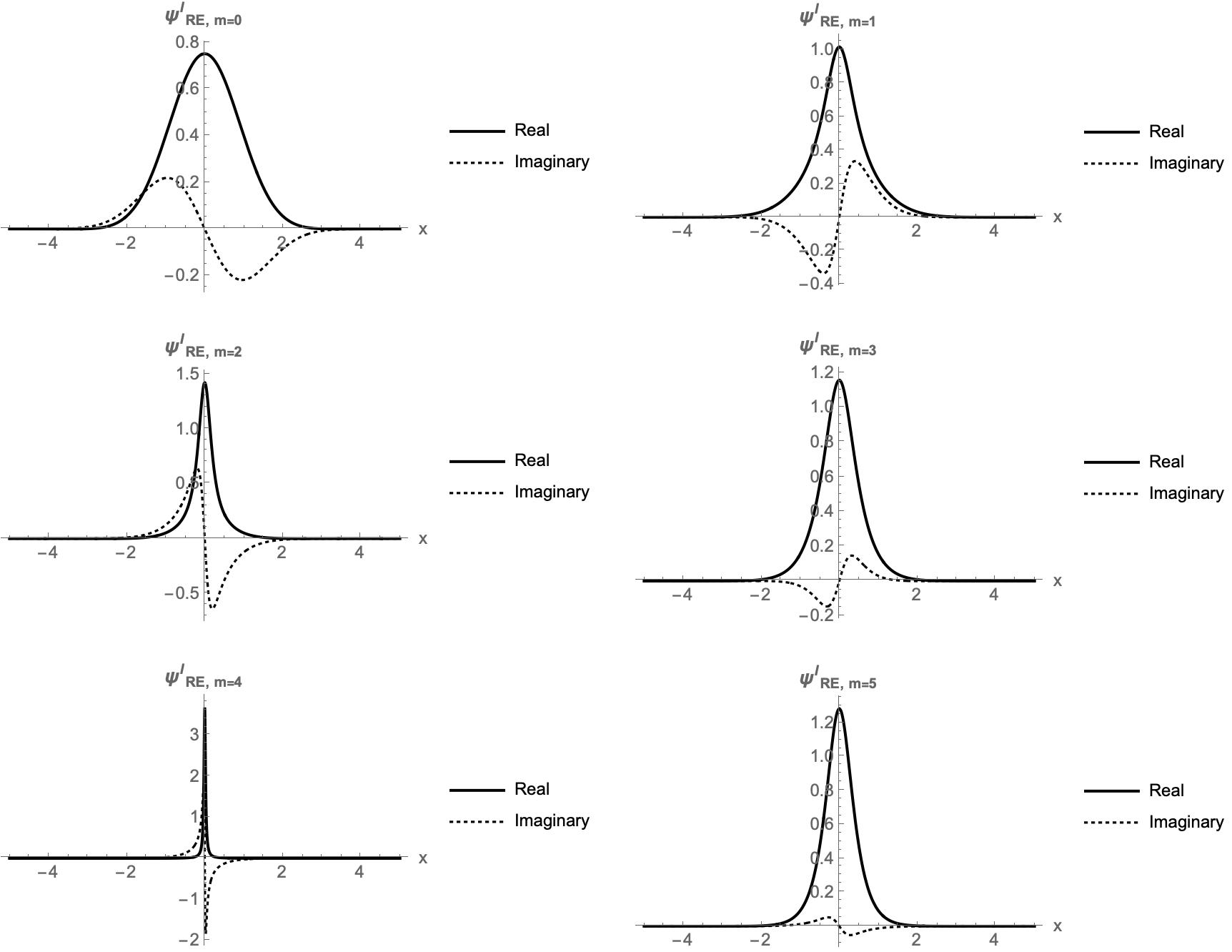}
  }{
      \fbox{\textbf{Figure Missing}}
  }
  \caption{Plots of real and imaginary components of $\psi^l_{RE,m,0}(x)$ vs $x$ for $m=0$ to $5$ and $\omega_1=2$.}
  \label{fig:1D-psi0-}
\end{figure}

Let us now discuss the case when the extended potential (\ref{1D-V-RE-I}) is 
complex but $\mathcal{PT}$ invariant by considering the case when $\lambda_0$
is pure imaginary, i.e. \(\lambda_0 = i\gamma_0\) with $\gamma_0$ real. 
Remarkably, the extended potential is now well-defined for even as well as odd 
\( m \), as can be checked from the Table \ref{Tab:1D-V-}. It is amusing to 
note that while only even co-dimension $m$ is allowed for the 
rational extension of the one dimensional QHO, both odd and 
even co-dimensions are allowed in the case of the rational extension of the one 
dimensional QHO plus an imaginary but $\mathcal{PT}$-invariant perturbation. 
All the energy eigenvalues are real and are independent of the parameter 
$\gamma_0$. Note that in this case the ground as well as the excited state 
eigenfunctions (\ref{1D-psi-RE-I}) are also eigenfunction of the 
$\mathcal{PT}$ operator i.e.

\begin{eqnarray}
		\mathcal{PT}\;\td{\psi}_{RE,m,0}^l(\td{x})&=&\td{\psi}_{RE,m,0}^l(\td{x})\nonumber\\
	\mbox{and} \quad  \mathcal{PT}\;\td{\psi}_{RE,m,n+1}^l(\td{x})&=&(-1)^{n+1}\;\td{\psi}_{RE,m,n+1}^l(\td{x})
\end{eqnarray}
respectively. The expressions for the ground and the excited state 
eigenfunctions for the first few values of \( m \) are also easily obtained 
from Table \ref{Tab:1D-psi0-} and Table \ref{Tab:1D-psin-} respectively after 
replacing $\lambda_0\rightarrow i\gamma_0$. 
The plots for the potential $V_{RE,m}^{l}(x)$ in case \(\lambda_0 = i\) 
(i.e. $\gamma_0=1$), are given in Fig.~\ref{fig:1D-V-} in case $\omega_1 = 2$ 
and $m = 0$ to $5$ while the corresponding ground state eigenfunction plots are
given in Fig.~\ref{fig:1D-psi0-}.


\begin{table}[htp]
  \centering
  \renewcommand{\arraystretch}{1.5} 
  \setlength{\tabcolsep}{1em} 
  \begin{tabularx}{\columnwidth}{|c|X|}
  \toprule
  {\bf m} & $\mathbf{V_{RE,m}^l(x)}$ \\
  \toprule
  0 & $\frac{\omega_1^2 x^2}{4} + \lambda_0 x - \omega_1$ \\
  \hline
  1 & $\frac{\omega_1^2 x^2}{4} + \lambda_0 x - \omega_1 + \frac{2 \omega_1^4}{(2 \lambda_0 + \omega_1^2 x)^2}$ \\
  \hline
  2 & $\frac{\omega_1^2 x^2}{4} + \lambda_0 x - \omega_1 - \frac{8 \omega_1^7}{\left(4 \lambda_0^2 + \omega_1^3 + \omega_1^4 x^2 + 4 \lambda_0 \omega_1^2 x\right)^2} + \frac{4 \omega_1^4}{4 \lambda_0^2 + \omega_1^3 + \omega_1^4 x^2 + 4 \lambda_0 \omega_1^2 x}$ \\
  \hline
  3 & $\frac{\omega_1^2 x^2}{4} + \lambda_0 x - \omega_1 - \frac{24 \omega_1^7}{\left(4 \lambda_0^2 + 3 \omega_1^3 + \omega_1^4 x^2 + 4 \lambda_0 \omega_1^2 x\right)^2} + \frac{4 \omega_1^4}{4 \lambda_0^2 + 3 \omega_1^3 + \omega_1^4 x^2 + 4 \lambda_0 \omega_1^2 x} + \frac{2 \omega_1^4}{(2 \lambda_0 + \omega_1^2 x)^2}$ \\
  \bottomrule
  \end{tabularx}
  \caption{RE potentials $V_{RE,m}^l(x)$ for different $m$ values in old coordinates.}
  \label{Tab:1D-V-}
\end{table}

\begin{table}[htp]
  \centering
  \renewcommand{\arraystretch}{2} 
  \setlength{\tabcolsep}{1em} 
  \begin{tabularx}{\textwidth}{|c|X|}
    \toprule
    {\bf m} & $\mathbf{\psi_{RE,m,0}^l(x)}$ \\
    \toprule
    0 & $\zeta\left(\omega_1,x + \frac{2\lambda_0}{\omega_1^2}\right)$ \\
    \midrule
    1 & $\zeta\left(\omega_1,x + \frac{2\lambda_0}{\omega_1^2}\right)\left[\frac{(-\omega_1)^{3/2}}{\sqrt{2} (2 \lambda_0 + x \omega_1^2)}\right]$ \\
    \midrule
    2 & $\zeta\left(\omega_1,x + \frac{2\lambda_0}{\omega_1^2}\right)\left[-\frac{\omega_1^3}{2 (4 \lambda_0^2 + 4 x \lambda_0 \omega_1^2 + \omega_1^3 + x^2 \omega_1^4)}\right]$ \\
    \midrule
    3 & $\zeta\left(\omega_1,x + \frac{2\lambda_0}{\omega_1^2}\right)\left[\frac{(-\omega_1)^{9/2}}{2 \sqrt{2} (2 \lambda_0 + x \omega_1^2) (4 \lambda_0^2 + 4 x \lambda_0 \omega_1^2 + \omega_1^3 (3 + x^2 \omega_1))}\right]$ \\
    \bottomrule
  \end{tabularx}
  \caption{Ground state eigenfunctions $\psi_{RE,m,0}^l(x)$ for different $m$ in old coordinates.}
  \label{Tab:1D-psi0-}
\end{table}

\begin{table}[htp]
  \centering
  \renewcommand{\arraystretch}{2.2} 
  \setlength{\tabcolsep}{0.5em} 
  \begin{tabularx}{\textwidth}{|c|X}
  \toprule
  {\bf m} & $\mathbf{\psi_{RE,m,n+1}^l(x)}$ \\
  \toprule
  0 & 
  $\zeta\left(\omega_1,x + \frac{2\lambda_0}{\omega_1^2}\right)\left[\frac{\left(2 \lambda_0 + \omega_1^2 x\right)}{\omega_1} H_n\left(\sqrt{\frac{\omega_1}{2}}\left(x + \frac{2 \lambda_0}{\omega_1^2}\right)\right) 
  - \sqrt{2} n \sqrt{\omega_1} H_{n-1}\left(\sqrt{\frac{\omega_1}{2}}\left(x + \frac{2 \lambda_0}{\omega_1^2}\right)\right)\right]$ \\
  \midrule
  1 & 
  $\scriptstyle\zeta\left(\omega_1,x + \frac{2\lambda_0}{\omega_1^2}\right)\left[\frac{\left(4 \lambda_0^2 + \omega_1^3 + \omega_1^4 x^2 + 4 \lambda_0 \omega_1^2 x\right)}{2 \lambda_0 \omega_1 + \omega_1^3 x} H_n\left(\sqrt{\frac{\omega_1}{2}}\left(x + \frac{2 \lambda_0}{\omega_1^2}\right)\right) 
  - \sqrt{2} n \omega_1^{3/2} \frac{\left(2 \lambda_0 + \omega_1^2 x\right)}{\omega_1} H_{n-1}\left(\sqrt{\frac{\omega_1}{2}}\left(x + \frac{2 \lambda_0}{\omega_1^2}\right)\right)\right]$ \\
  \midrule
  2 & 
  $\scriptstyle\zeta\left(\omega_1,x + \frac{2\lambda_0}{\omega_1^2}\right)\left[\frac{\left(8 \lambda_0^3 + 6 \lambda_0 \omega_1^3 (\omega_1 x^2 + 1) + \omega_1^5 x (\omega_1 x^2 + 3) + 12 (\lambda_0^2) \omega_1^2 x\right)}{4 \lambda_0^2 \omega_1 + \omega_1^4 + \omega_1^5 x^2 + 4 \lambda_0 \omega_1^3 x} H_n\left(\sqrt{\frac{\omega_1}{2}}\left(x + \frac{2 \lambda_0}{\omega_1^2}\right)\right) 
  - \sqrt{2} n \sqrt{\omega_1} H_{n-1}\left(\sqrt{\frac{\omega_1}{2}}\left(x + \frac{2 \lambda_0}{\omega_1^2}\right)\right)\right]$ \\
  \midrule
  3 & 
  $\zeta\left(\omega_1,x + \frac{2\lambda_0}{\omega_1^2}\right)\frac{\left(16 \lambda_0^4 + 24 (\lambda_0^2) \omega_1^3 (\omega_1 x^2 + 1) + 8 \lambda_0 \omega_1^5 x (\omega_1 x^2 + 3) + \omega_1^6 (\omega_1^2 x^4 + 6 \omega_1 x^2 + 3) + 32 \lambda_0^3 \omega_1^2 x\right)}{\omega_1 \left(2 \lambda_0 + \omega_1^2 x\right) \left(4 \lambda_0^2 + \omega_1^3 (\omega_1 x^2 + 3) + 4 \lambda_0 \omega_1^2 x\right)} H_n\left(\sqrt{\frac{\omega_1}{2}}\left(x + \frac{2 \lambda_0}{\omega_1^2}\right)\right)$ \\
  \bottomrule
  \end{tabularx}
  \caption{Excited state eigenfunctions $\psi_{RE,m,n+1}^l(x)$ for different $m$ in old coordinates.}
  \label{Tab:1D-psin-}
\end{table}

\newpage

\section{Two-dimensional QAHO potential with quadratic perturbation}

In this section, we consider a 2D-quantum anisotropic oscillator potential with
a quadratic perturbation term $\frac{\lambda}{2} xy$ (with $\lambda$ being real
or imaginary) and angular frequencies \( \omega_1 \) and \( \omega_2 \) along 
the \( x \) and \( y \) axes, respectively, given by
\begin{equation}\label{2D-V-II}
 V(x,y)= \frac{1}{4}\omega_1^2x^2+\frac{1}{4}\omega_2^2y^2+\frac{\lambda}{2} xy.
\end{equation}
The transformation from the old to the new coordinates and vice versa are given
by \cite{mandal2013pt}
\begin{align}\label{2D-xy}
  \begin{split}
  \begin{aligned}
      x &= a\td{x}+b\td{y}, \\
      y &= -b\td{x}+a\td{y},
  \end{aligned}
  & 
  \left\{
  \begin{aligned}
      \td{x} &= ax-by, \\
      \td{y} &=bx+ay,
  \end{aligned}
  \right.
  \end{split}
\end{align}
where $a=\sqrt{\frac{1-k}{2}}$, $b=\sqrt{\frac{1+k}{2}}$ and \( k =\frac{\omega_1^2-\omega_2^2}{\sqrt{4\lambda^2+(\omega_1^2-\omega_2^2)^2}}\). 
On using the above co-ordinate transformations, the resulting 
uncoupled potential in the new co-ordinates, $\tilde{V}(\td{x},\td{y})$ turns 
out to be 
\begin{equation}\label{2D-V-II'}
\tilde{V}(\td{x},\td{y})= \td{V}(\td{x})+\td{V}(\td{y}),  
\end{equation}
where, $\td{V}(\td{x}),\;\td{V}(\td{y})$ are the QHO potentials in the new 
coordinates, i.e.
\begin{equation}\label{2D-V-degeneracy}
  \td{V}(\td{x},\td{y})= \frac{1}{4}\left(\td{\omega}_1^2\td{x}^2+\td{\omega}_2^2\td{y}^2\right).
\end{equation}
with the new angular frequencies $\td{\omega}_1$ and $\td{\omega}_2$ given by
\begin{align}\label{2D-omega'}
  \begin{split}
    \td{\omega}_1 &= \sqrt{\frac{1}{2} \left(\omega_1^2+\omega_2^2-\sqrt{4\lambda^2+(\omega_1^2-\omega_2^2)^2}\right)}\\
 \mbox{and} \quad \td{\omega}_2 &= \sqrt{\frac{1}{2} \left(\omega_1^2+\omega_2^2+\sqrt{4\lambda^2+(\omega_1^2-\omega_2^2)^2}\right)}
  \end{split}
\end{align}
respectively. The corresponding eigenfunctions $\td{\psi}_{n_1,n_2}(\td{x},\td{y})$ and the eigenvalues $\td{E}_{n_1,n_2}(\td{\omega}_1,\td{\omega}_2)$ 
are given by
\begin{align}\label{2D-psi-II'}
  \tilde{\psi}_{n_1,n_2}(\td{x},\td{y}) &\propto
  \td{\psi}_{n_1}(\td{x})\;\td{\psi}_{n_2}(\td{y}),\quad n_1,n_2 = 0,1,2,\cdots, \\\label{2D-E-II'}
  \mbox{and}\quad \tilde{E}_{n_1,n_2}(\td{\omega}_1,\td{\omega}_2) &= \left(n_1+\frac{1}{2}\right)\td{\omega}_1+\left(n_2+\frac{1}{2}\right)\td{\omega}_2,
\end{align}
respectively, where $\td{\psi}_{n}(\td{x})$ is the eigenfunction of the 1D-QHO 
as given by Eq. (\ref{QHO-psi}) but in new coordinates with new angular 
frequencies defined by (\ref{2D-omega'}).
Depending on whether $\lambda$ is real or imaginary, the spectrum is real 
within specific ranges.We discuss these two cases one by one.

{\bf Case(a) Hermitian case:} When $\lambda$ is real, the conditions for real spectra is
\begin{equation}
  \sqrt{4 \lambda ^2+(\omega _1^2-\omega_2^2)^2}\leq\omega_1^2+\omega_2^2,\label{real2D}
\end{equation}
which implies a restriction on $\lambda$ i.e., 
$|\lambda| \leq \omega_1 \omega_2$.

{\bf Case(b) Non-hermitian and $\mathcal{PT}$ symmetric case:} When $\lambda$ 
is imaginary, say $\lambda=i\gamma$, the system is non-hermitian and 
$\mathcal{PT}$ symmetric and the spectrum is real only when the following 
conditions are satisfied \cite{mandal2013pt} 
\begin{align}\label{img2D}
  \begin{split}
    \sqrt{-4\gamma ^2+(\omega _1^2-\omega _2^2)^2}&\le\omega _1^2+\omega _2^2,\\
   \mbox{and}\quad -4 \gamma ^2+(\omega _1^2-\omega _2^2)^2&>0.
  \end{split}
\end{align}
These conditions imply a restriction on $\gamma$ i.e., 
$|\gamma|< \frac{1}{2} |\omega _1^2-\omega _2^2|$.
The $\eta$-pseudo hermiticity for imaginary $\lambda$ is not identity and is 
given by \cite{mostafazadeh2002pseudo1} 
\begin{equation}
  \eta=\left(
\begin{array}{cc}
 -k & -\sqrt{1-k^2}  \\
 \sqrt{1-k^2}  & -k \\
\end{array}
\right)
\end{equation}
and the potential $V(x,y)$ satisfies the $\eta$-pseudo hermiticity condition
\begin{equation}
  V^{\dagger}=\eta V \eta^{-1}.
\end{equation}
It is worth pointing out that in two space dimensions, parity transformation 
corresponds to either $(x,y) \rightarrow (-x,y)$ or $(x,y) \rightarrow (x,-y)$.
Thus the system is $\mathcal{PT}$ symmetric for the parity operators $P_1$ and 
$P_2$ given in (Eq. \ref{2D-P}) and the eigenfunction, $\tilde{\psi}_{n_1,n_2}(\td{x},\td{y})$, satisfies the following relations
\begin{align*}
P_1T\tilde{\psi}_{n_1,n_2}(\td{x},\td{y}) &= (-1)^{n_1}\tilde{\psi}_{n_1,n_2}(\td{x},\td{y}) = \pm\tilde{\psi}_{n_1,n_2}(\td{x},\td{y}), \\
P_2T\tilde{\psi}_{n_1,n_2}(\td{x},\td{y})&= (-1)^{n_2}\tilde{\psi}_{n_1,n_2}(\td{x},\td{y}) = \pm\tilde{\psi}_{n_1,n_2}(\td{x},\td{y})\,.
\end{align*}

\subsection{Rational Extension}
It is worth pointing out that the rational extension of the quantum anisotropic harmonic 
oscillator (QAHO) potentials in two dimensions can be performed in four 
distinct ways \cite{Kumar:2024tsz}, leading to four different forms of the 
rationally extended potentials. These possibilities depend on whether both $x$ 
and $y$ are on the full line, or both on the half line, and one on the full 
line while the other is on the half line. However in this paper we consider 
the case when both $x$ and $y$ are defined on the full lines.

As shown in our recent work \cite{Kumar:2024tsz}, the rational extension of 
the QAHO potential in two and higher dimensions is simply the sum of the 
rational extensions of the QAHO potentials along each direction. It turns out 
that only even co-dimensions are possible as the new co-ordinates \(\tilde{x}\) 
and \(\tilde{y}\) are simply the linear functions of the old coordinates \(x\) 
and \(y\). Thus, the rationally extended potential corresponding to the 
potential (\ref{2D-V-II'}) for even co-dimensions $m_1$ and $m_2$ is given by 
\begin{align}\label{2D-V-RE-II}
  \td{V}_{RE,m_1,m_2}(\td{x},\td{y}) &= \td{V}_{RE,m_1}(\td{x})+\td{V}_{RE,m_2}(\td{y}),
\end{align}
where $\td{V}_{RE,m_1}(\td{x})$ and $\td{V}_{RE,m_2}(\td{y})$ are the rational 
extension of the QHO along the $\td{x}$ and the $\td{y}$ directions. The
form of these potentials is the same as that of the 1D case given by 
Eq. (\ref{1D-V-RE-x}). Doing inverse coordinate transformations, we get the 
expression for the rationally extended potential $V_{RE,m_1,m_2}(x,y)$ in the 
old coordinates but in terms of the new frequencies $\td{\omega}_1$ and 
$\td{\omega}_2$. In Table-\ref{Tab:2D-V-} we have given the expressions for 
$V_{RE,m_1,m_2}(x,y)$ for few values of \( m_1 \) and \( m_2 \).
The corresponding ground and the excited state eigenfunctions and the 
energy eigenvalues are given Using Eqs. (\ref{1D-psi-RE-x}) and 
(\ref{1D-E_RE-x}) in the new coordinates as 
\begin{align}\label{2D-psi-RE-II}
	\td{\psi}_{RE,m_1,m_2,0,0}(\td{x},\td{y}) &\propto \td{\psi}_{RE,m_1,0}(\td{x})\;\td{\psi}_{RE,m_2,0}(\td{y})\\
  \td{\psi}_{RE,m_1,m_2,n_1+1,n_2+1}(x,y) &\propto \td{\psi}_{RE,m_1,n_1+1}(\td{x})\;\td{\psi}_{RE,m_2,n_2+1}(\td{y})\\
  \begin{split}
 \mbox{and} \quad \td{E}_{RE,m_1,m_2,n_1+1,n_2+1}(\td{\omega}_1,\td{\omega}_2) &=\td{E}_{RE,m_1,n_1+1}(\td{\omega}_1)+\td{E}_{RE,m_2,n_2+1}(\td{\omega}_2)
\end{split}
\end{align}
\begin{equation}
\mbox{with} \quad \td{E}_{RE,m_1,m_2,0,0}(\td{\omega}_{1},\td{\omega}_{2}) =0, 
\quad n_1, n_2=0,1,2... \,. \nonumber
\end{equation}
The rationally extended potential $\td{V}_{RE,m_1,m_2}(\td{x},\td{y})$  is 
nonsingular for even $m$ irrespective of $\lambda$ being real or imaginary. In 
the real $\lambda$ case, the spectrum will be real only when the condition 
(\ref{real2D}) is satisfied. In the case of imaginary $\lambda$ 
(say $\lambda=i\gamma$), the potential 
$\td{V}_{RE,m_1,m_2}(\td{x},\td{y})$ is $\mathcal{PT}$ symmetric under 
both the parity operators $P\rightarrow P_1$ or $P_2$ (\ref{2D-P}) and the 
spectrum is real when the condition (\ref{img2D}) is satisfied. 
The ground and the excited state eigenfunctions corresponding to this RE 
potential also satisfy
 \begin{align*}
 PT\;\td{\psi}_{RE,m_1,m_2,0,0}(\td{x},\td{y}) &= \td{\psi}_{RE,m_1,m_2,0,0}(\td{x},\td{y})\\
PT\;\td{\psi}_{m_1,m_2,n_1+1,n_2+1}(\td{x},\td{y})&= (-1)^{n_1+n_2+2}\td{\psi}_{m_1,m_2,n_1,n_2}(\td{x},\td{y}) = \pm\td{\psi}_{m_1,m_2,n_1,n_2}(\td{x},\td{y})\,.
 \end{align*}    

\subsection{Conditions for Degeneracy}
The system will show degeneracy when the ratio of the angular frequencies in 
the new coordinates is a rational number. i.e.,
\begin{equation}
  \frac{\td{\omega}_1}{\td{\omega}_2}=\td{r}\,.
\end{equation}
Substituting the values of $\td{\omega}_1$ and $\td{\omega}_2$ from 
Eq. (\ref{2D-omega'}) and simplifying for $\lambda$ we get
\begin{equation}\label{2D-lambda}
  \lambda(\td{r},r,\omega_1,\omega_2)=\frac{\sqrt{(\td{r}^4+1)-\left(\frac{\td{r}}{r}\right)^2(r^4+1)}}{\td{r}^2+1}\;\omega_1\omega_2\,.
\end{equation}
Here $r = \frac{\omega_1}{\omega_2}$, which may be a rational or irrational 
number. The energy eigenvalues of the potential (\ref{2D-V-degeneracy}) are 
given by
\begin{equation}\label{2D-degeneracy}
 \td{E}_{n_1,n_2}(\td{r},\td{\omega}_2) = \left(\td{r} n_1+  n_2+\frac{\td{r}+1}{2}\right)\td{\omega}_2.
\end{equation}

As an illustration we now consider two explicit examples, one when $\lambda$ is
real and one when it is pure imaginary.

\subsubsection{Example of Degeneracy in Case $\lambda$ is Real}



{\bf Rational Angular Frequencies \((\omega_1, \omega_2, \lambda) = \left(1, 2, \frac{\sqrt{7}}{2}\right)\)}\\

Thus the original potential has the form
\begin{equation}\label{img-example-2D}
  V(x,y)= \frac{1}{4}\left(\frac{1}{2}x^2+\frac{9}{2}y^2\right)+ \frac{\sqrt{7}}{4} xy.
\end{equation}
Doing coordinate transformations as discussed above and using  $k=-\frac{3}{4}$ gives
\[\tilde{V}(\td{x},\td{y})=\frac{1}{4} \left(\frac{1}{2} \td{x}^2+\frac{9}{2}\td{y}^2\right),\]
and the angular frequencies in new coordinates have ratio 
$\td{\omega}_1:\td{\omega}_2=1:3$ and the system is degenerate. It is now 
straight forward to obtain the corresponding rationally extended potential 
$V_{RE,m_1,m_2}(x,y)$. As an illustration the corresponding rationally extended 
potentials \( V_{RE,0,2}(x, y) \) for \( m_1, m_2 = 0, 2 \) and 
\( m_1, m_2 = 2, 2 \) are respectively
\begin{align*}
  V_{RE,0,2}(x, y) &=V(x,y) -\frac{1}{\sqrt{2}} \\
  &\quad - 2 \left( \frac{3}{2\sqrt{2}} 
  - \frac{72 \left( \frac{x}{2\sqrt{2}} + \frac{1}{2} \sqrt{\frac{7}{2}} y \right)^2}
         {\left( -2 - 3\sqrt{2} \left( \frac{x}{2\sqrt{2}} + \frac{1}{2} \sqrt{\frac{7}{2}} y \right)^2 \right)^2} 
  - \frac{6\sqrt{2}}
         {-2 - 3\sqrt{2} \left( \frac{x}{2\sqrt{2}} + \frac{1}{2} \sqrt{\frac{7}{2}} y \right)^2} \right)
\end{align*}

and
\begin{align*}
  V_{RE,2,2}(x, y) &= V(x,y)-2 \left( \frac{1}{2\sqrt{2}} 
  - \frac{8 \left( \frac{1}{2} \sqrt{\frac{7}{2}} x - \frac{y}{2\sqrt{2}} \right)^2}
         {\left( -2 - \sqrt{2} \left( \frac{1}{2} \sqrt{\frac{7}{2}} x - \frac{y}{2\sqrt{2}} \right)^2 \right)^2} 
  - \frac{2\sqrt{2}}
         {-2 - \sqrt{2} \left( \frac{1}{2} \sqrt{\frac{7}{2}} x - \frac{y}{2\sqrt{2}} \right)^2} \right) \\
  &\quad - 2 \left( \frac{3}{2\sqrt{2}} 
  - \frac{72 \left( \frac{x}{2\sqrt{2}} + \frac{1}{2} \sqrt{\frac{7}{2}} y \right)^2}
         {\left( -2 - 3\sqrt{2} \left( \frac{x}{2\sqrt{2}} + \frac{1}{2} \sqrt{\frac{7}{2}} y \right)^2 \right)^2} 
  - \frac{6\sqrt{2}}
         {-2 - 3\sqrt{2} \left( \frac{x}{2\sqrt{2}} + \frac{1}{2} \sqrt{\frac{7}{2}} y \right)^2} \right)
\end{align*}

\begin{figure}[htp]
  \centering
  \begin{minipage}{0.45\textwidth}
    \centering
    \IfFileExists{2D-V-02.jpeg}{
      \includegraphics[width=\textwidth]{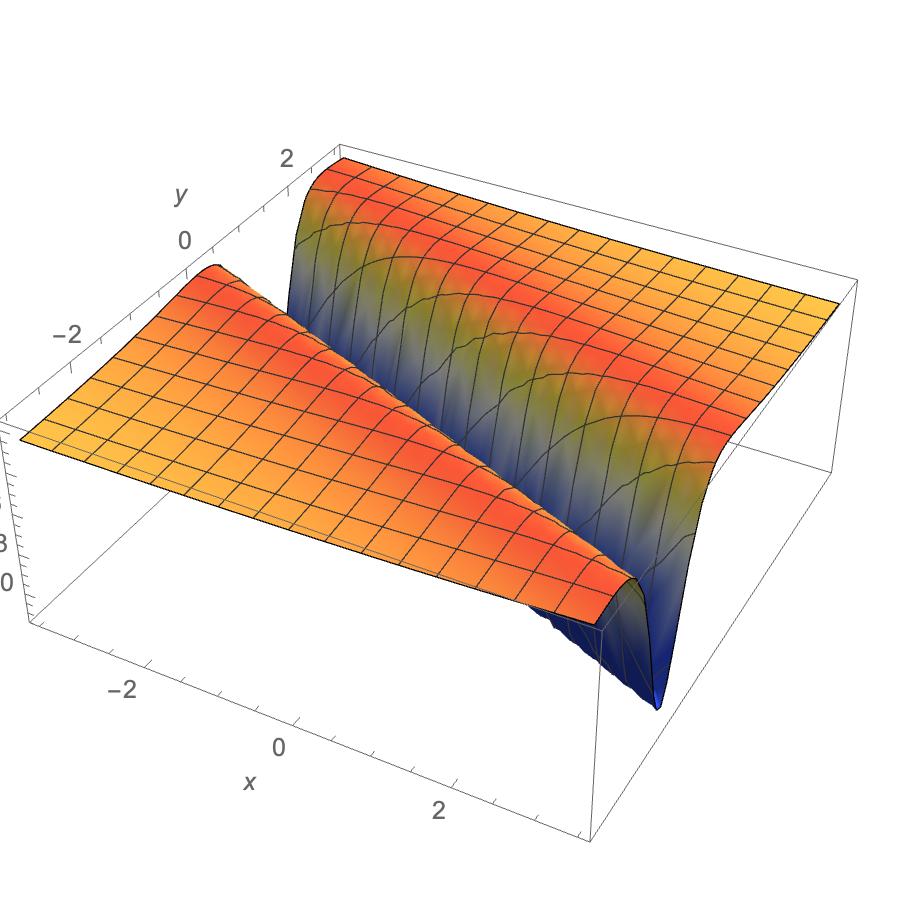}
    }{
      \fbox{\textbf{Figure Missing}}
    }
    \caption*{$V_{RE,,m_1=0,m_2=2}(x,y)$}
  \end{minipage}
  \hfill
  \begin{minipage}{0.45\textwidth}
    \centering
    \IfFileExists{2D-V-22.jpeg}{
      \includegraphics[width=\textwidth]{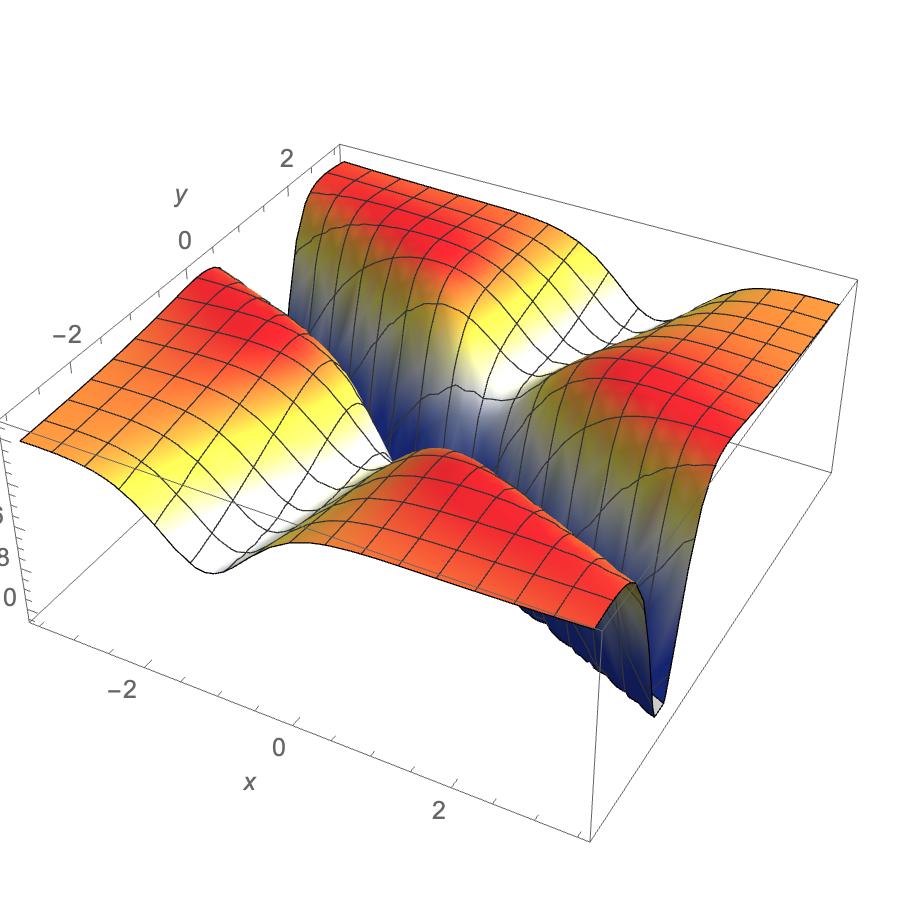}
    }{
      \fbox{\textbf{Figure Missing}}
    }
    \caption*{$V_{RE,m_1=2,m_2=2}(x,y)$}
  \end{minipage}
  \hfill
  \begin{minipage}{0.45\textwidth}
    \centering
    \IfFileExists{2D-V-24.jpeg}{
      \includegraphics[width=\textwidth]{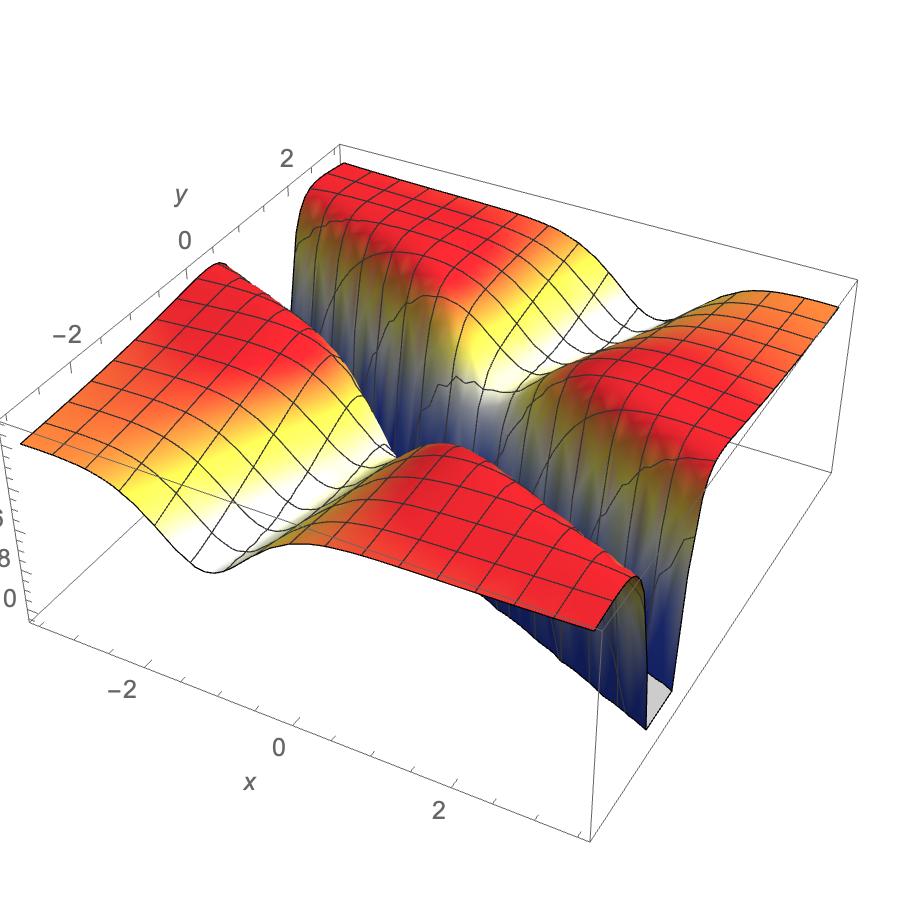}
    }{
      \fbox{\textbf{Figure Missing}}
    }
    \caption*{$V_{RE,m_1=2,m_2=4}(x,y)$}
  \end{minipage}
  \hfill
  \begin{minipage}{0.45\textwidth}
    \centering
    \IfFileExists{2D-V-44.jpeg}{
      \includegraphics[width=\textwidth]{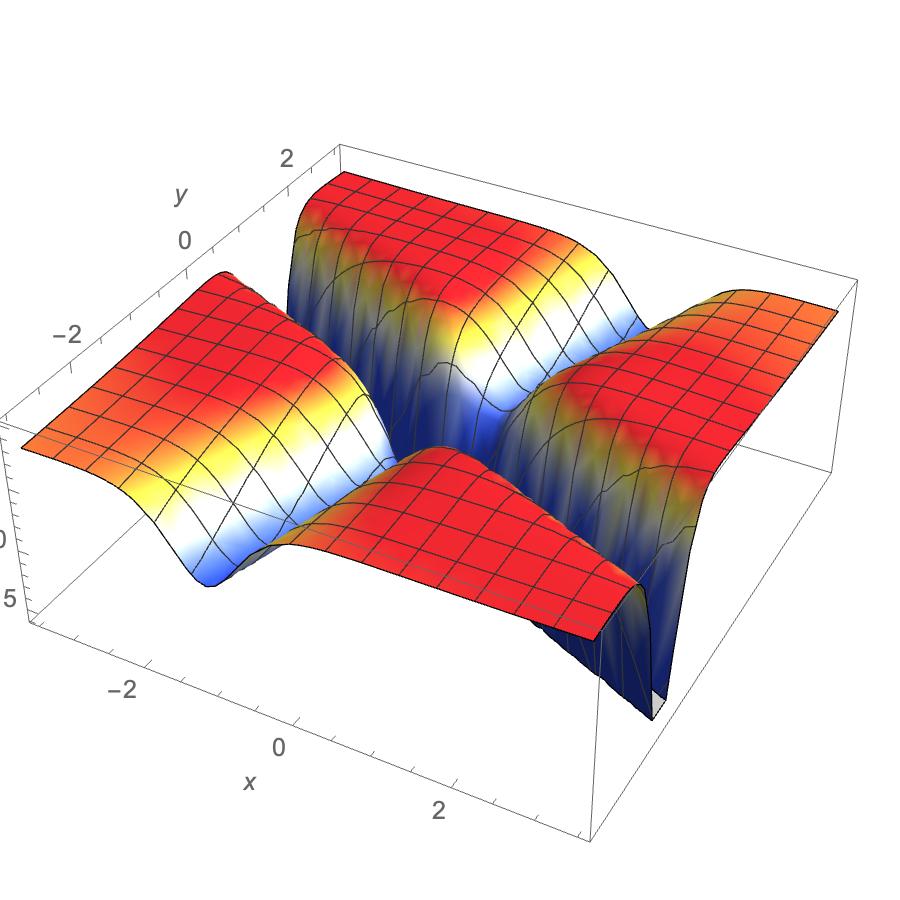}
    }{
      \fbox{\textbf{Figure Missing}}
    }
    \caption*{$V_{RE,m_1=4,m_2=4}(x,y)$}
  \end{minipage}
  \caption{Plots of the rationally extended potentials \( V_{RE,m_1,m_2}(x,y) \) as functions of \( x \) and \( y \) for different values of \( m_1 \) and \( m_2 \).}
  \label{fig:2D-V-re}
\end{figure}

The plot of the potential $V_{RE,m_1,m_2}(x,y)$ is shown in 
Fig.~\ref{fig:2D-V-re} in case $m_1 = 0, m_2 = 2; m_1 = 2, m_2 = 2; m_1 = 2, 
m_2 = 4$ and $m_1 = 4, m_2 = 4$. \\

\subsubsection{Example of Degeneracy in Case $\lambda$ is Imaginary}

\begin{figure}[htp] 
  \centering
  \begin{minipage}{0.45\textwidth}
    \centering
    \IfFileExists{2D-V-re02.jpeg}{
      \includegraphics[width=\textwidth]{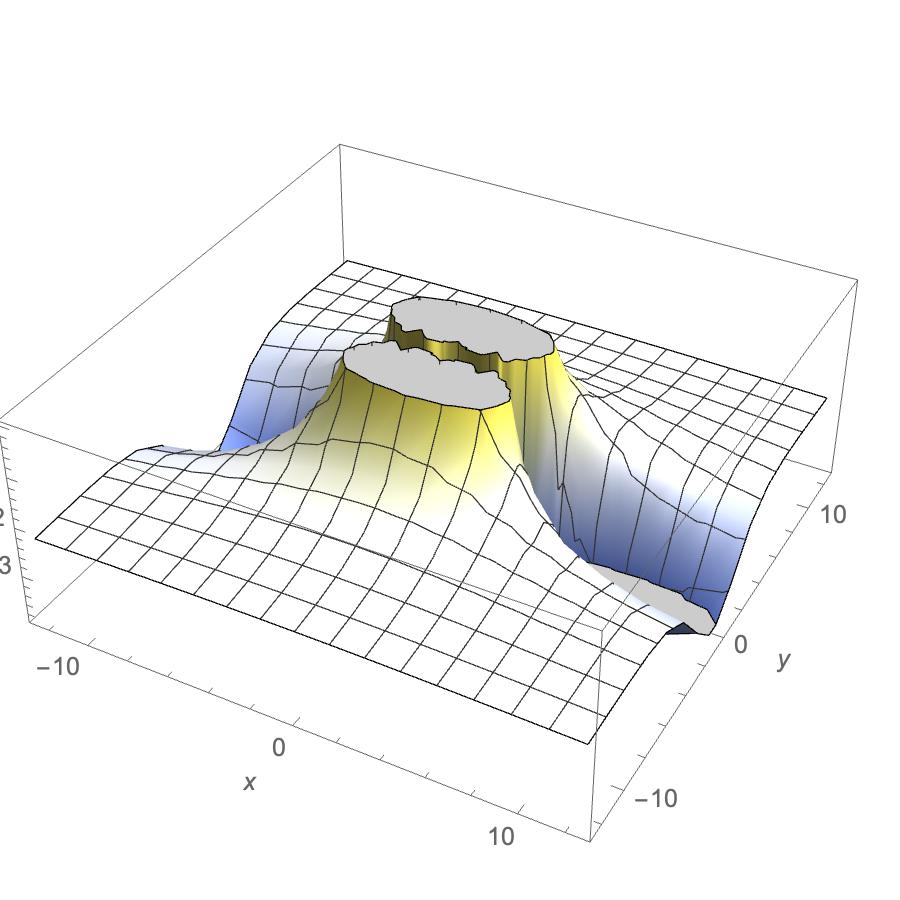}
    }{
      \fbox{\textbf{Figure Missing}}
    }
    \caption*{$Re[V_{RE,m_1=0,m_2=2}(x,y)]$}
  \end{minipage}
  \hfill
  \begin{minipage}{0.45\textwidth}
    \centering
    \IfFileExists{2D-V-img02.jpeg}{
      \includegraphics[width=\textwidth]{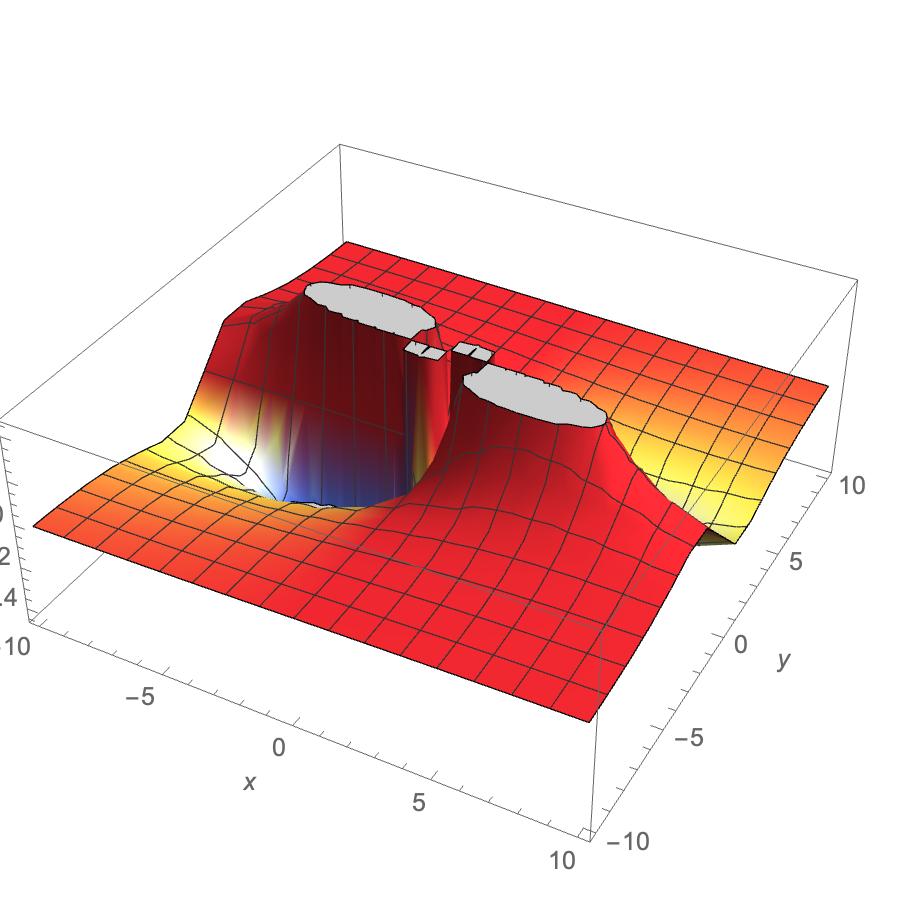}
    }{
      \fbox{\textbf{Figure Missing}}
    }
    \caption*{$Im[V_{RE,m_1=0,m_2=2}(x,y)]$}
  \end{minipage}
  \hfill
  \begin{minipage}{0.45\textwidth}
    \centering
    \IfFileExists{2D-V-re22.jpeg}{
      \includegraphics[width=\textwidth]{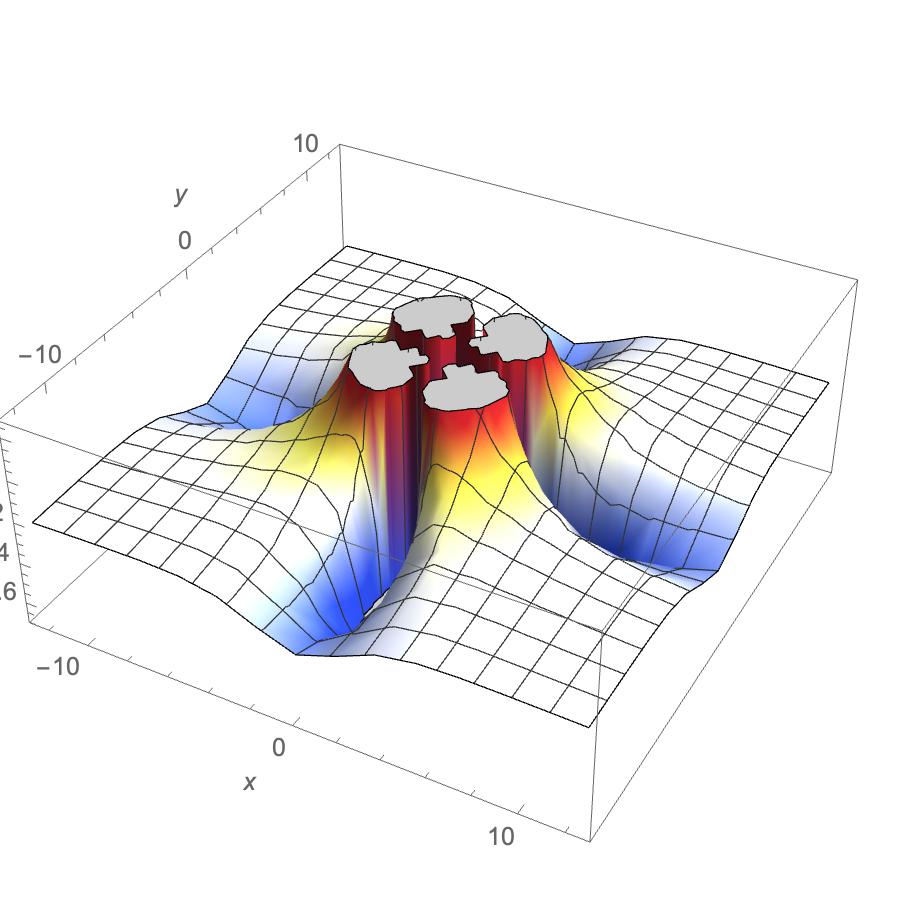}
    }{
      \fbox{\textbf{Figure Missing}}
    }
    \caption*{$Re[V_{RE,m_1=2,m_2=2}(x,y)]$}
  \end{minipage}
  \hfill
  \begin{minipage}{0.45\textwidth}
    \centering
    \IfFileExists{2D-V-img22.jpeg}{
      \includegraphics[width=\textwidth]{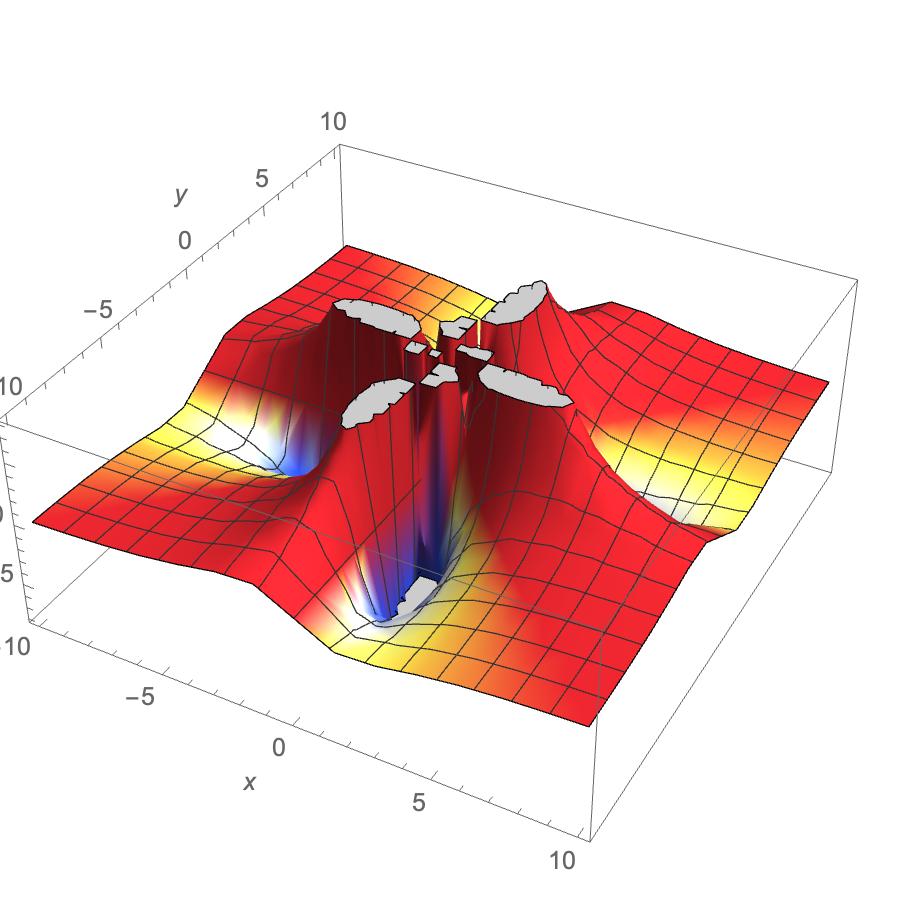}
    }{
      \fbox{\textbf{Figure Missing}}
    }
    \caption*{$Im[V_{RE,m_1=2,m_2=2}(x,y)]$}
  \end{minipage}
  \caption{Plots of \( V_{RE, m_1, m_2}(x, y) \) vs. \( x \) and \( y \), showing the Real part (left) and Imaginary part (right) for various values of \( m_1 \) and \( m_2 \), with imaginary \( \lambda \).}
  \label{fig:2D-V-img}
\end{figure}

As an illustration, we consider  \((\omega_1, \omega_2, \lambda) 
= \left(1, 3, i\sqrt7\right)\), so that the  original potential has the form
\begin{equation}\label{img-example-2D}
  V(x,y)= \frac{1}{4}\left(x^2+9y^2\right)+ i\frac{\sqrt{7}}{2} xy.
\end{equation}
Using the transformation given by Eq. (\ref{2D-xy}) with $k=-\frac{4}{3}$, one 
obtains
\[\tilde{V}(\td{x},\td{y})=\frac{1}{4} \left(2 \td{x}^2+8\td{y}^2\right)\]
and therefore the ratio of the angular frequencies in the new coordinates is 
$\td{\omega}_1:\td{\omega}_2=1:2$ and the system is degenerate. The 
corresponding rationally extended potential for arbitrary even $m_1,m_2$ is
easily calculated. As an illustration, the corresponding rationally 
extended potential in case $m_1=m_2=2$ is
\begin{align*}
  \begin{split}
  V_{RE,2,2}(x,y) &= V(x,y)-2 \left(-\frac{128 \left(\sqrt{\frac{7}{6}} y+\frac{i x}{\sqrt{6}}\right)^2}{\left(-2-4 \sqrt{2} \left(\sqrt{\frac{7}{6}} y+\frac{i x}{\sqrt{6}}\right)^2\right)^2}-\frac{8 \sqrt{2}}{-2-4 \sqrt{2} \left(\sqrt{\frac{7}{6}} y+\frac{i x}{\sqrt{6}}\right)^2}+\sqrt{2}\right)\\&-2 \left(-\frac{32 \left(\sqrt{\frac{7}{6}} x-\frac{i y}{\sqrt{6}}\right)^2}{\left(-2-2 \sqrt{2} \left(\sqrt{\frac{7}{6}} x-\frac{i y}{\sqrt{6}}\right)^2\right)^2}-\frac{4 \sqrt{2}}{-2-2 \sqrt{2} \left(\sqrt{\frac{7}{6}} x-\frac{i y}{\sqrt{6}}\right)^2}+\frac{1}{\sqrt{2}}\right).
\end{split}
\end{align*}
The plot for the real and imaginary part of the potential $V_{RE,m_1,m_2}(x,y)$
are given in Fig. \ref{fig:2D-V-img} in case $m_1 = 0, m_2 =2$ and 
$m_1 = m_2 = 2$.

Before ending this section, it is worth pointing out that if we consider 
anisotropic oscillator in two dimensions with the perturbation of the form
$i(\lambda_1 x+ \lambda_2 y)$ then following the treatment of the last section,
we can easily obtain the corresponding rational potential for odd as well 
as even co-dimensions $m_1$ and $m_2$ since the problems in the $x$ and $y$ 
coordinates essentially decouple.

 

\begin{table}[htp]
  \centering
  \renewcommand{\arraystretch}{1.5} 
  \setlength{\tabcolsep}{1em} 
  \begin{tabularx}{\columnwidth}{|c|X|}
    \toprule
    $\mathbf{m_1,m_2}$ & $\mathbf{V_{RE,m_1,m_2}(x,y)}$ \\
    \toprule
    $\mathbf{0,0}$ &$ V(x,y)-\td{\omega}_1-\td{\omega}_2 $ \\
    \hline
    $\mathbf{0,2}$ &$V(x,y)-\td{\omega}_1-\td{\omega}_2+\frac{4 \td{\omega}_2 \left(\td{\omega}_2 (a y+b x)^2-1\right)}{\left(\td{\omega}_2 (a y+b x)^2+1\right)^2} $\\
    \hline
    $\mathbf{2,2}$ &$V(x,y)-\td{\omega}_1-\td{\omega}_2 +\frac{4 \td{\omega}_2 \left(\td{\omega}_2 (a y+b x)^2-1\right)}{\left(\td{\omega}_2 (a y+b x)^2+1\right)^2}+\frac{4 \td{\omega}_1 \left(\td{\omega}_1(a x - b y)^2 -1\right)}{\left((a x - b y)^2 \td{\omega}_1+1\right)^2}$ \\
    \hline
    $\mathbf{2,4}$ &$V(x,y)-\td{\omega}_1-\td{\omega}_2$\\
      & $+\frac{4 \left(2 - (b (x - 2 y) + a (2 x + y)) (a (-2 x + y) + b (x + 2 y)) \td{\omega}_1 + (a x - b y)^2 \left(-2 a b x y + a^2 (2 x^2 + y^2) + b^2 (x^2 + 2 y^2)\right) \td{\omega}_1^2\right)}{\left(1 + (a x - b y)^2 \td{\omega}_1\right)^2}$\\
      & $+\frac{4 \left(42 - 18 (b x + a y)^2 \td{\omega}_2\right)}{ \left(3 + (b x + a y)^2 \td{\omega}_2 \left(6 + (b x + a y)^2 \td{\omega}_2\right)\right)}+\frac{576 \left(-1 - 2 (b x + a y)^2 \td{\omega}_2\right)}{\left(3 b x + 3 a y + 6 (b x + a y)^3 \td{\omega}_2 + (b x + a y)^5 \td{\omega}_2^2\right)^2}$\\
    \bottomrule
    \end{tabularx}  
    \caption{Potentials $V_{RE,m_1,m_2}(x,y)$ for even $(m_1,m_2)$ values in old coordinates.}
\label{Tab:2D-V-}
\end{table}

\newpage

\section{Three-dimensional QAHO potentials}
In this section, we consider a three-dimensional QAHO potential (defined on the 
full-line)  along with the following two different perturbations 

1. Combination of a linear and a quadratic perturbation  
$(\lambda_0z+\frac{\lambda}{2}xy)$ 

2. Quadratic perturbations $(\lambda_1 xy +\lambda_2 yz +\lambda_3 zx)$. 

In each case, we will decouple the perturbed potential using an appropriate 
co-ordinate transformation and then obtain their rational extension. 

\subsection{Combination of a linear and a quadratic perturbations}
In this subsection, we consider a three-dimensional QAHO potential 
$V^{lq}(x,y,z)$ with linear and  quadratic perturbations, given by
\begin{equation}\label{3D-V-III}
V^{lq}(x,y,z) = \frac{1}{4} \left(\omega_{1}^2 x^2 + \omega_{2}^2 y^2 
+ \omega_{3}^2 z^2\right) + \lambda_{0} z + \frac{\lambda}{2} xy.
\end{equation}
which can be re-expressed as a mixture of one-dimensional potential with a 
linear perturbation $V^l(z)$ and two-dimensional anisotropic potential with a 
quadratic perturbation $V(x,y)$ as 
\begin{equation}
  V^{lq}(x, y, z) =  V(x, y)+ V^l(z),
\end{equation}
which can be further decoupled in the tilde co-ordinates using the 
transformations (\ref{1D-V'}) and (\ref{2D-V-II'}) as
\begin{equation}
\td{V}^{lq}(\td{x}, \td{y}, \td{z}) =  \td{V}(\td{x}, \td{y})+ \td{V}^l(\td{z}).
\end{equation}
 The corresponding eigenfunctions can be easily obtained by using Eqs.  
 (\ref{QHO-psi}) and (\ref{2D-psi-II'}) 
\begin{equation}\label{lq_ef}
  \td{\psi}^{lq}_{n_1, n_2, n_3}(\td{x}, \td{y}, \td{z}) \propto \td{\psi}_{n_1, n_2}(\td{x}, \td{y}) \td{\psi}^l_{n_3}( \td{z})
\end{equation}
and the corresponding energy eigenvalues are given by the sum of the energy 
eigenvalues of the 2D and the 1D cases as
\begin{equation}\label{lq_en}
  \td{E}^{lq}_{n_1, n_2, n_3}(\td{\omega}_1, \td{\omega}_2, \td{\omega}_3) = \td{E}_{n_1, n_2}(\td{\omega}_1, \td{\omega}_2) + \td{E}_{n_3}(\omega_3)
\end{equation}

We need to consider the four possible cases depending on if $\lambda_0$ and
$\lambda$ are real or purely imaginary.

\begin{enumerate}

\item {\bf Both parameters $\lambda_0$ and $\lambda$ are real}

In this case the energy spectrum will be real only when the equation 
(\ref{real2D}) is satisfied. Further, in this case the rational extension is
possible only if all three co-dimensions $m_1, m_2, m_3$ are even.

  \item {\bf $\lambda_0$ is imaginary  but $\lambda$ is real}
  
 The system is $\mathcal{PT}$-invariant where the parity operator 
		$\mathcal{P}_2$ [see Eq. (\ref{3D-P})] is 
 \begin{equation}
	 \mathcal{P}_2 : x \rightarrow x,\, y \rightarrow y,\, z \rightarrow -z\,. 
 \end{equation}
In this case the $\mathcal{PT}$-symmetry is unbroken and the energy 
eigenvalues (\ref{lq_en}) are real for all the values of $\gamma$ 
(note $\lambda_0 = i\gamma$ with $\gamma$ real) and the corresponding 
eigenfunctions (\ref{lq_ef}) are $\mathcal{PT}$-invariant, i.e.
  \begin{equation}
  	\mathcal{P}_2 \mathcal{T}\; \td{\psi}^{lq}_{n_1, n_2, n_3}(\td{x}, \td{y}, \td{z})=(-1)^{n_3} \td{\psi}^{lq}_{n_1, n_2, n_3}(\td{x}, \td{y}, \td{z}).\nonumber
  \end{equation}
  
  \item {\bf When $\lambda_0$ is real while $\lambda$ is imaginary}
  
In this case, the system is $\mathcal{PT}$-invariant 
where the parity operators are either $\mathcal{P}_1$ or $\mathcal{P}_3$  
[see Eq. (\ref{3D-P})] i.e.,
  \begin{equation}
  \mathcal{P}_1 : x \rightarrow -x,\, y \rightarrow y,\, z \rightarrow z\,, \\ 
\mathcal{P}_3 : x \rightarrow x,\, y \rightarrow -y,\, z \rightarrow z\,. 
 \end{equation}
As discussed in the previous section, the spectrum will be real when the 
conditions (\ref{img2D}) are satisfied and the eigenfunctions (\ref{lq_ef}) 
satisfy
  \begin{equation}
  	\mathcal{P}_1 \mathcal{T}\; \td{\psi}^{lq}_{n_1, n_2, n_3}(\td{x}, \td{y}, \td{z})=\mathcal{P}_3 \mathcal{T} \psi^{lq}_{n_1, n_2, n_3}(\td{x}, \td{y}, \td{z})=(-1)^{n_1+n_2} \td{\psi}^{lq}_{n_1, n_2, n_3}(\td{x}, \td{y}, \td{z}).\nonumber	
  \end{equation} 
 
 \item {\bf When both $\lambda_0$ and $\lambda$ are imaginary}
 
 In this case, the potential is $\mathcal{PT}$-invariant where the parity 
operator $\mathcal{P}_4$ [see Eq. (\ref{3D-P})] is
 \begin{equation}
  \mathcal{P}_4 : x \rightarrow -x,\, y \rightarrow -y,\, z \rightarrow -z\,.
 \end{equation}
 and the eigenfunctions satisfy
 \begin{equation}
 \mathcal{P}_4 T\; \td{\psi}^{lq}_{n_1, n_2, n_3}(\td{x}, \td{y}, \td{z})=(-1)^{n_1+n_2+n_3}\td{\psi}^{lq}_{n_1, n_2, n_3}(\td{x}, \td{y}, \td{z}). \nonumber
 \end{equation} 
\end{enumerate}

\subsubsection{Rational Extension of $\td{V}^{lq}(\td{x}, \td{y}, \td{z})$}

The rational extension of the real as well as all the $\mathcal{PT}$-symmetric 
cases of the potential $\td{V}^{lq}(\td{x}, \td{y}, \td{z})$ in the tilde co-ordinates is simply the sum of the rational extensions of the 2D- QAHO and the 1D QHO potentials given by Eqs. (\ref{2D-V-RE-II}) and  (\ref{1D-V-RE-I}) respectively as
\begin{align}
  \begin{split}
    \td{V}^{lq}_{RE,m_1,m_2,m_3}(\td{x}, \td{y}, \td{z}) &= \td{V}_{RE,m_1,m_2}(\td{x}, \td{y})+\td{V}^l_{RE,m_3}(\td{z})
  \end{split}
\end{align}
where $\td{V}_{RE,m_1,m_2}(\td{x}, \td{y})$ and $\td{V}^l_{RE,m_3}(\td{z})$ are
the rationally extended potentials corresponding to the potentials 
$\td{V}(\td{x},\td{y})$ and $\td{V}^l(\td{z})$ as given by 
Eqs. (\ref{2D-V-II'}) and (\ref{1D-V'}) respectively. To get the above extended
potential in terms of the old co-ordinates, we use the inverse co-ordinate 
transformations. Thus, the ground and the excited state eigenfunctions with 
their corresponding energy eigenvalues are given by
\begin{align}
	\td{\psi}^{lq}_{RE,m_1,m_2,m_3,0,0,0}(\td{x}, \td{y}, \td{z})&\propto \td{\psi}_{RE,m_1,m_2,0,0}(\td{x}, \td{y})\;\td{\psi}^l_{RE,m_3,0}(\td{z}),
\end{align}
\begin{align}
  \td{\psi}^{lq}_{RE,m_1,m_2,m_3,n_1+1,n_2+1,n_3+1}(\td{x}, \td{y}, \td{z})&\propto \td{\psi}_{RE,m_1,m_2,n_1+1,n_2+1}(\td{x}, \td{y})\;\td{\psi}^l_{RE,m_3,n_3+1}(\td{z})
\end{align}
and 
\begin{align}
  \td{E}^{lq}_{RE,m_1,m_2,m_3,n_1+1,n_2+1,n_3+1}(\td{\omega}_1,\td{\omega}_2,\td{\omega}_3)&=\td{E}_{RE,m_1,m_2,n_1+1,n_2+1}(\td{\omega}_1,\td{\omega}_2)+\td{E}^l_{RE,m_3,n_3+1}(\td{\omega}_3)
\end{align}
with
\begin{equation}
	\td{E}^{lq}_{RE,m_1,m_2,m_3,0,0,0}(\td{\omega}_1,\td{\omega}_2,\td{\omega}_3)=\td{E}_{RE,m_1,m_2,0,0}(\td{\omega}_1,\td{\omega}_2)+\td{E}^l_{RE,m_3,0}(\td{\omega}_3)\nonumber
\end{equation}
respectively. Here $n_1,n_2,n_3=0,1,2,.....,$. We now briefly discuss the 
allowed co-dimensions in the four possible cases as discussed above.

\begin{itemize}
\item {\bf When both $\lambda$ and $\lambda_0$ are real}: 

In this case, the potential is regular only for the even values of the 
co-dimension $m_1,m_2$ and $m_3$.

\item{\bf When $\lambda$ is real and $\lambda_0$ is imaginary}:

In this case, the potential is  regular only for the even values of 
the co-dimension $m_1,m_2$ but for both even and odd co-dimension $m_3$.

\item{\bf When $\lambda$ is imaginary and $\lambda_0$ is real}:

In this case,  the potential is regular only for the even values of the 
co-dimensions $m_1,m_2$ and $m_3$.

\item{\bf When both $\lambda$ and $\lambda_0$ are imaginary}:
	
In this case the potential is regular only for the even values of the 
co-dimension $m_1,m_2$ but for both even and odd co-dimension $m_3$.

\end{itemize}


\subsection{3D QAHO Potential with Combinations of Quadratic Perturbations}

Now, we consider another example of a three-dimensional QAHO potential with a 
combination of quadratic perturbations given by
\begin{equation}
V(x,y,z) = \frac{1}{4}\left(\omega_1^2x^2 + \omega_2^2y^2 + \omega_3^2z^2 + 2\lambda_1 xy + 2\lambda_2 yz + 2\lambda_3 zx\right). \label{3D-V}
\end{equation}
Even though the perturbed potential can, in principle, be decoupled into the 
standard form of QHO by applying a co-ordinate transformation, however, in 
practice, for a general perturbation of the form 
\(\lambda_1 xy +\lambda_2 yz +\lambda_3 zx\), it is not easy to find a general 
co-ordinate transformation with arbitrary frequencies and the coupling parameters that will reduce the 
potential to a pure anisotropic oscillator in the new coordinates. We therefore
discuss here one special case when two of the three frequencies are equal. 

As an illustration, first we consider the case \(\omega_1 = \omega_2=\omega\;\text{(say)}\), in which the potential in (\ref{3D-V}) simplifies to
\begin{equation}
V^{q}(x,z,y) = \frac{1}{4}\left(\;\omega^2(x^2 + y^2) + \omega_3^2z^2 + 2\lambda_1 xy + 2\lambda_2 yz + 2\lambda_3 zx\right). \label{special-3D}
\end{equation}
Further, we assume two cases: either \(\lambda_1 = 0\) with \(\lambda_2\ne\lambda_3\), and \(\lambda_1 \neq 0\) with \(\lambda_2 = \lambda_3\), which are discussed in detail in the following subsections: 

\subsubsection{ Case I: $\mathbf{\lambda_1=0}$ and $\mathbf{\lambda_2\ne \lambda_3}$}
In this case, the potential (\ref{special-3D}) is given by
\begin{equation}
	V^{q_1}(x,y,z)=\frac{1}{4}\left(\omega^2(x^2+y^2)+\omega_3^2z^2+2\lambda_2 yz+2\lambda_3 zx\right)\label{vq}
\end{equation}
To solve this problem, we use the following transformations from 
the old to the new coordinates and vice-versa 
\begin{align}\label{3D-xyz-IV}
  \begin{split}
  \begin{aligned}
      x &= a d \td{y} + bd\td{z} - c\td{x}, \\
      y &= d\; \td{x} + ac\td{y} + bc\td{z} \\
      z &=a\td{z} - b\td{y}, 
  \end{aligned}
  & 
  \left\{
  \begin{aligned}
    \td{x} &= - cx + d\;y  , \\
    \td{y} &=ad\;x + acy - bz, \\
    \td{z} &= bd\;x + bcy + az.
  \end{aligned}
  \right.
  \end{split}
\end{align}
where \(k = \frac{\omega^2 - \omega_3^2}{\sqrt{4 \lambda_2^2 + 4 \lambda_3^2 
+ \left(\omega^2 - \omega_3^2\right)^2}}\), 
$c=\frac{\lambda_2}{\sqrt{\lambda_2^2+\lambda_3^2}}$ and 
$d=\frac{\lambda_3}{\sqrt{\lambda_2^2+\lambda_3^2}}$. The other two parameters 
$a$ and $b$ are same as defined in the 2D case by Eq. (\ref{2D-xy}). Thus, the 
potential in the new co-ordinates is given by 
\begin{equation}
\tilde{V}^{q_1}(\td{x},\td{y},\td{z})=\frac{1}{4}\left(\td{\omega}_1^2\td{x}^2+\td{\omega}_2^2 \td{y}^2+\td{\omega}_3^2\td{z}^2\right)	\label{V_q}
\end{equation}
where the new frequencies are given by
\begin{align}\label{3D-omega'-IV}
  \begin{split}
    \td{\omega}_1 &= \omega, \\
    \td{\omega}_2 &= \sqrt{\frac{1}{2}\left(\omega^2+\omega_3^2 - \sqrt{4(\lambda_2^2 + \lambda_3^2) + \left(\omega^2-\omega_3^2\right)^2}\right)}, \\
    \mbox{and} \quad \td{\omega}_3 &= \sqrt{\frac{1}{2}\left(\omega^2+\omega_3^2 + \sqrt{4(\lambda_2^2 + \lambda_3^2) + \left(\omega^2-\omega_3^2\right)^2}\right)}.
  \end{split}
\end{align}
\subsubsection*{Rational Extension:}

Similar to the 2D case, to get the rational extension of the above potential 
(\ref{vq}), we use the starting potential (\ref{V_q}) and get the corresponding
RE potential in the new co-ordinates which is transformed back to  the old 
co-ordinates with perturbation terms using the inverse co-ordinate 
transformation (\ref{3D-xyz-IV}). The expression of the RE potential in the new
co-ordinate is given by 
\begin{align}
	\td{V}^{q_1}_{RE,m_1,m_2,m_3}(\td{x},\td{y},\td{z})= \td{V}_{RE,m_1}(\td{x})+\td{V}_{RE,m_2}(\td{y})+\td{V}_{RE,m_3}(\td{z})
\end{align}
where $(\td{V}_{RE,m_1}(\td{x}),\;\td{V}_{RE,m_2}(\td{y}),\;
\td{V}_{RE,m_3}(\td{z}))$ are given by Eq. (\ref{1D-V-RE-x}) but in terms of 
the new angular frequencies and the new coordinates. Similarly the ground and 
the excited state eigenfunctions are written using rationally extended 
eigenfunctions in the new coordinates and the new angular frequencies as
\begin{align}\nonumber
	\td{\psi}^{q_1}_{RE,m_1,m_2,m_3,0,0,0}(\td{x},\td{y},\td{z})\propto \td{\psi}_{RE,m_1,0}(\td{x})\;
  \td{\psi}_{RE,m_2,0}(\td{y})\; \td{\psi}_{RE,m_3,0}(\td{z})
\end{align}
and 
\begin{align}
	\td{\psi}^{q_1}_{RE,m_1,m_2,m_3,n_1+1,n_2+1,n_3+1}(\td{x},\td{y},\td{z})\propto \td{\psi}_{RE,m_1,n_1+1}(\td{x})\;
  \td{\psi}_{RE,m_2,n_2+1}(\td{y})\; \td{\psi}_{RE,m_3,n_3+1}(\td{z})
\end{align}
respectively. The corresponding energy eigenvalues in terms of the new angular 
frequencies are given by
\begin{align}\nonumber
	\td{E}^{q_1}_{RE,m_1,m_2,m_3,n_1+1,n_2+1,n_3+1}(\td{\omega}_1,\td{\omega}_2,\td{\omega}_3)&=\td{E}_{RE,m_1,n_1+1}\left(\td{\omega}_1\right)\\
	&+\td{E}_{RE,m_2,n_2+1}\left(\td{\omega}_2\right)+\td{E}_{RE,m_3,n_3+1}\left(\td{\omega}_3\right)
\end{align}
with
\begin{align}
	\td{E}^{q_1}_{RE,m_1,m_2,m_3,0,0,0}(\td{\omega}_1,\td{\omega}_2,\td{\omega}_3)&=0
\end{align}
The conditions for the real spectra for the various choices of $\lambda$ are as
follows:
\begin{enumerate}
  \item {\bf Hermitian case:} When $\lambda$'s are real, then real spectra are 
obtained in case 
  \begin{equation}\label{cond-3D-r-1}
      0 \leq \lambda_2^2 + \lambda_3^2 \leq \omega^2 \omega_3^2
  \end{equation}   

  \item {\bf Non-Hermitian and $\mathcal{PT}$-symmetric case:} With imaginary 
$\lambda$'s, the system shows $\eta$-pseudo-hermiticity and hence the system is
$\mathcal{PT}$- symmetric for the different parity operators depending on 
whether one or both the coupling parameters ($\lambda_2$ and $\lambda_3$) are 
imaginary. Thus, we have three different cases as discussed below:  
  
    {\bf (a) When $\lambda_3$ is real and $\lambda_2$ is imaginary (say $\lambda_2=i\gamma_2$)}\\ 	
    The system exhibits \(\mathcal{PT}\) symmetry with the parity operator 
\(\mathcal{P}_{3}\) (see Eq. (\ref{3D-P})), defined as 
    \[\mathcal{P}_{3} : x \rightarrow x, \quad y \rightarrow -y, \quad z 
\rightarrow z.\] 
and the real spectra is obtained in this case when \\
    \[-\frac{1}{4}(\omega^2-\omega_3^2)^2 \leq -\gamma_2^2 + \lambda_3^2 \leq \omega^2 \omega_3^2.\]
    
    {\bf (b) When $\lambda_3$ is imaginary (say $\lambda_3=i\gamma_3$) and $\lambda_2$ is real}\\ 
     Here \(\mathcal{PT}\) symmetry is defined by the parity operator 
\(\mathcal{P}_{1}\) (see Eq. (\ref{3D-P})), where 
    \[\mathcal{P}_{1} : x \rightarrow -x, \quad y \rightarrow y, \quad z \rightarrow z.\]
    and the spectrum will be real when

    \[-\frac{1}{4}(\omega^2-\omega_3^2)^2 \leq \lambda_2^2 - \gamma_3^2 \leq \omega^2 \omega_3^2.\]

  { \bf (c) When both $\lambda$'s are imaginary (say, $\lambda_2=i\gamma_2$ and $\lambda_3=i\gamma_3$)}\\
 In this case the system is $\mathcal{PT}$ symmetric for the parity operator \(\mathcal{P}_{2}\) (Eq. (\ref{3D-P}))
  	\begin{align*}
  		\mathcal{P}_2& : x \rightarrow x, \quad y \rightarrow y, \quad z \rightarrow -z,
  	\end{align*}
  and the real spectrum is obtained in case  
  
  \[ \gamma_2^2 + \gamma_3^2\leq\frac{1}{4}(\omega^2-\omega_3^2)^2.\]
   \end{enumerate}

\subsubsection*{Condition for Degeneracy:}
The condition for the degeneracy in the system is similar to the 2D case and is
given by
\begin{equation}
  \sqrt{\lambda_2^2 + \lambda_3^2} =\frac{\sqrt{(\td{u}^4+1)-\left(\frac{\td{u}}{u}\right)^2(u^4+1)}}{\td{u}^2+1}\;\omega\omega_3
\end{equation}
where $\td{u}=\frac{\td{\omega}_2}{\td{\omega}_3}$ is a rational number and $u=\frac{\omega}{\omega_3}$ can be rational or irrational.

As an illustration, we consider the case when $\omega=\sqrt 2$ and 
$ \omega_3=1$, so that $u=\sqrt2$. In this case, the potential takes the form 
\[V(x,y,z)=2 x^2+2 y^2+z^2+\lambda_2 y z+\lambda_3 x z\].
Now, if we consider \(\td{u} = 2\), we get $\lambda^2_2+\lambda^2_3=\frac{14}{25}$, which gives various combinations  of \((\lambda_2, \lambda_3)\) such as 
\[
(\lambda_2, \lambda_3) = \left(\frac{\sqrt{7}}{5}, \frac{\sqrt{7}}{5}\right), \left(\frac{\sqrt6}{5}, \frac{\sqrt{8}}{5}\right), \left(0, \frac{\sqrt{14}}{5}\right), \cdots
\]
leading to the degeneracy.

\subsubsection{Case II:  $\mathbf{\lambda_1\ne0}$ and $\mathbf{\lambda_2=\lambda_3}$}
In case $\lambda_2=\lambda_3=\lambda\text{(say)}$, with $\lambda$ being real or
imaginary, the potential (\ref{3D-V}) is given by
\begin{equation}
V^{q_2}(x,y,z)=\frac{1}{4}\left[\omega^2(x^2+y^2)+\omega_z^2z^2
+2\lambda_1xy+2\lambda(yz+ zx)\right].
\end{equation}
The transformation relations from the old to the new coordinates and vice-versa
are given by,
\begin{align}\label{3D-xyz-V}
  \begin{split}
  \begin{aligned}
      x &= \frac{a}{\sqrt2}\td{y}+\frac{b}{\sqrt2} \td{z}-\frac{\td{x}}{\sqrt{2}}, \\
      y &= \frac{a}{\sqrt2} \td{y}+\frac{b}{\sqrt2}\td{z}+\frac{\td{x}}{\sqrt{2}},\\
      z &=a\td{z}-b\td{y}, 
  \end{aligned}
  & 
  \left\{
  \begin{aligned}
      \td{x} &= \frac{y}{\sqrt{2}}-\frac{x}{\sqrt{2}} , \\
      \td{y} &=\frac{a}{\sqrt2} x+\frac{a}{\sqrt2}y-bz , \\
      \td{z} &= \frac{b}{\sqrt2} x+\frac{b}{\sqrt2}y+az.
  \end{aligned}
  \right.
  \end{split}
\end{align}
where \(k = \frac{ \omega^2 - \omega_3^2+q\lambda_1}{\sqrt{8 \lambda^2 + \left(\omega^2 - \omega_3^2+q\lambda_1\right)^2}}\), and the new frequencies are 
\begin{align}
  \label{3D-omega'-V}
  \begin{split}
    \td{\omega}_1 &= \sqrt{\omega^2 -  \lambda_1}, \\
    \td{\omega}_2 &= \sqrt{\frac{1}{2}\left(\omega^2+\omega_3^2 +  \lambda_1 - \sqrt{8\lambda^2 + (\omega^2-\omega_3^2 +  \lambda_1)^2}\right)}, \\
    \td{\omega}_3 &= \sqrt{\frac{1}{2}\left(\omega^2+\omega_3^2 +  \lambda_1 + \sqrt{8\lambda^2 + (\omega^2-\omega_3^2 +  \lambda_1)^2}\right)}.
  \end{split}
\end{align}

\subsubsection*{Rational Extension}
The rational extension in this case is similar to the previous case except the 
new frequencies and new coordinates are given by Eqs. (\ref{3D-omega'-V}) and  
(\ref{3D-xyz-V}) respectively. 

{\bf The condition on $\lambda$'s for real spectra:} 

From the expressions for the new frequencies as given by 
Eq. (\ref{3D-omega'-V}), it is clear that while the parameter \(\lambda_1\) 
has to be real, the parameter $\lambda$ can be either real or imaginary. This 
distinction gives rise to the two possibilities for the 
real spectra which are discussed below:

\begin{enumerate}
\item {\bf Hermitian case:} When $\lambda$ is real, then the spectra is real
provided
      \begin{equation}\label{cond-3D-r-2}
        \begin{split}
            -\omega^2\le\lambda_1\le \omega^2\qquad\qquad\\
           \mbox{and} \quad 0\leq \lambda_1^2\leq \frac{(\omega^2+\lambda)\omega_z^2}{2}.
        \end{split}  
      \end{equation}

 \item {\bf Non-Hermitian and $\mathcal{PT}$ symmetric case:} When $\lambda$ is
imaginary, say $\lambda=i \gamma$, the allowed values of $\gamma^2, \lambda_1$ 
and $\omega$ are 
      \begin{equation}\label{cond-3D-i-2}
            -\omega^2\le\lambda_1\le\omega^2 \quad
            \mbox{and}\quad 0\leq \gamma^2\leq \frac{(\omega^2-\omega_3^2+\lambda_1)^2}{8}
        \end{equation}  
      The system is $\mathcal{PT}$-invariant with the corresponding parity 
operator $\mathcal{P}_2$ (see eq. (\ref{3D-P})) i.e.,
      \begin{align*}
\mathcal{P}_2& : x \rightarrow x, \quad y \rightarrow y, \quad z \rightarrow -z.
      \end{align*}
  \end{enumerate}

\subsubsection*{Degeneracy Conditions:}

The condition for the degeneracy in this case is given by
\begin{equation}
  \sqrt{8\lambda^2+\lambda_1^2}=\frac{|\td{u}^2-1|}{\td{u}^2+1}\;(2\omega^2+\lambda_1),
\end{equation}
which simplifies to 
\begin{equation}
  \lambda=\frac{\sqrt{\left(\frac{|\td{u}^2-1|}{\td{u}^2+1}(\lambda_1+2\omega^2)\right)^2-\lambda_1^2}}{2 \sqrt{2}}.
\end{equation}
Whenever this condition is satisfied for some rational values of $\td{u}$, the 
system will exhibit degeneracy. For example, when 
\(\lambda = \frac{\omega^2}{4}\sqrt{\frac{15}{2}}\) and 
\(\lambda_1 = \frac{\omega^2}{2}\), so that \(\td{u} =\frac{1}{3}\), the system exhibits degeneracy.

Before ending this section, it is worth pointing out that if we consider 
anisotropic oscillator in three dimensions with the perturbation of the form
$i(\lambda_1 x+ \lambda_2 y+\lambda_3 z)$ then following the treatment of the
section II, we can easily obtain the corresponding rational potential for odd 
as well as even co-dimensions $m_1, m_2$ and $m_3$ since the problems in the 
$x, y$ and $z$ coordinates essentially decouple.

\section{Summary and Possible Open Problems}

In this work, we have obtained the rational extension of the quantum 
anisotropic oscillator (QAO) potentials under various perturbations in one, two
and three dimensions. First, we considered the case of $1D$ QHO with a linear 
perturbation and shown that unlike the pure harmonic oscillator (where the 
rational extension is only allowed for even co-dimension $m$), the rational 
extension exists for both even and odd co-dimensions  \(m\)  when the linear
perturbation is purely imaginary. However, if the linear perturbation is real
then the rational extension is only possible for even co-dimension $m$. 
In the two dimensional QAO case we obtained rational extension with either real
or imaginary quadratic perturbations in case the co-dimensions $m_1$ and $m_2$ 
are even. Further, we obtained the conditions for the spectrum to be real and 
related it with the unbroken PT-symmetry. Finally, we also considered the case 
of the $3D$ QAO potentials with combined linear and quadratic perturbation or
sum of pure quadratic perturbations. In both the cases we obtained the rational
extension in case the perturbing terms are real or pure imaginary or 
admixture of both.

This paper raises few obvious questions. For example, what are the various 
perturbations for which rational extensions are possible for the QAO in higher 
dimensions? Further, what are the various perturbations possible for which 
rational extension is possible in the case of the spherically symmetric 
oscillator potentials? Note that without the perturbation, as has been shown in 
\cite{non_cent} the rational extension is in terms of the exceptional Jacobi  
polynomials. We hope to address some of these issues in the near future.

\section*{Acknowledgement}
RKY acknowledges S.K.M. University, Dumka, for the grant sanctioned by 
University letter No. SKMU/CCDC/349, under the DHTE State Research Project for Teachers of State Universities, Jharkhand State 
Higher Education Council, Govt. of Jharkhand (India). AK is grateful to Indian National 
Science Academy (INSA) for awarding INSA Honorary Scientist position at 
Savitribai Phule Pune University.


\renewcommand{\theequation}{A.\arabic{equation}} 
\setcounter{equation}{0}


\renewcommand{\theequation}{A.\arabic{equation}} 
\setcounter{equation}{0} %

\section*{Appendix A: The concept of \( \mathcal{PT} \)-Symmetry in higher 
dimensions}

The parity ($\mathcal{P}$) and the time reversal ($\mathcal T$) transformations
in 1-D are defined as
\begin{align*}
\mathcal{P}:&x\rightarrow-x,\;i\rightarrow i\\
\mathcal{T}:&x\rightarrow x,\;i\rightarrow -i
\end{align*}
In 2-D, the parity transformation corresponds to either say $(\mathcal{P}_1)$
and $(\mathcal{P}_2)$, as \cite{mandal2013pt}  
\begin{align*}
\mathcal{P}_1:&x=-x,\quad y=y\\
\mathcal{P}_2:&x=x,\quad y=-y
\end{align*}
A matrix representation of \( \mathcal{P} \) (in any number of dimensions) has 
a determinant equal to \(-1\), distinguishing it from a rotation, which has a 
determinant equal to \(+1\). In matrix form, the above two parity operators are
given by
\begin{equation}\label{2D-P}
\mathcal{P}_1=\begin{bmatrix}
-1&0\\
0&1
\end{bmatrix};\quad\mbox{and}\quad
\mathcal{P}_2=\begin{bmatrix}
1&0\\
0&-1
\end{bmatrix}
\end{equation}
Clearly, there are other possible transformations all satisfying the condition 
that the determinant of the transformation matrix is -1. This suggests that the
parity operator is not unique and can be defined in many ways. For example, two
more possibilities are
\begin{equation}
\mathcal{P}_3=\begin{bmatrix}
0&1\\
1&0
\end{bmatrix};\quad \text{and }\quad
\mathcal{P}_4=\begin{bmatrix}
0&-1\\
-1&0
\end{bmatrix}
\end{equation}
Similarly, in 3D case, few possible parity transformation matrices are 
\begin{equation}\label{3D-P}
    \begin{aligned}
        \mathcal{P}_1 &= \begin{bmatrix}
            -1 & 0 & 0 \\
            0 & 1 & 0 \\
            0 & 0 & 1    
        \end{bmatrix} ; \quad &
        \mathcal{P}_2 &= \begin{bmatrix}
            1 & 0 & 0 \\
            0 & 1 & 0 \\
            0 & 0 & -1
        \end{bmatrix} ; \\
        \mathcal{P}_3 &= \begin{bmatrix}
            1 & 0 & 0 \\
            0 & -1 & 0 \\
            0 & 0 & 1
        \end{bmatrix} ; \quad &
        \mathcal{P}_4 &= \begin{bmatrix}
            -1 & 0 & 0 \\
            0 & -1 & 0 \\
            0 & 0 & -1
        \end{bmatrix} .
    \end{aligned}
\end{equation}
It can be seen that the first operator is defining the reflection about the 
plane \(x=0\), the second operator is defining the reflection about the plane 
\(z=0\), the third operator is defining the reflection about the plane \(y=0\),
and the fourth operator is defining the space inversion.

\end{document}